\documentclass[10pt]{article}
\usepackage{epsfig,amssymb,amsmath,psfrag}
\usepackage[T1]{fontenc}
\usepackage{psfrag}
\usepackage{geometry}
\numberwithin{equation}{section}

\catcode`\@=11
\def \be  {\begin{equation}}
\def \ee  {\end{equation}}
\def \ba  {\begin{eqnarray}}
\def \ea  {\end{eqnarray}}
\def \baa {\begin{eqnarray*}}
\def \eaa {\end{eqnarray*}}
\def \bbib  {\begin {thebibliography} }
\def \ebib  {\end{thebibliography}}
\def \lab #1 {\label{#1}}

\def \e  {\mathop{\rm e}\nolimits}
\newcommand{\ft}[2]{{\textstyle\frac{#1}{#2}}}

\newcommand{\mbf}[1]{\mbox{\boldmath $#1$}}
\newcommand{\bk}{\mbf{k}}
\newcommand{\bp}{\mbf{p}}
\newcommand{\bq}{\mbf{q}}
\newcommand{\bx}{\mbf{x}}
\newcommand{\bb}{\mbf{b}}

\begin{document}
\bibliographystyle{unsrt}

\title{Regge limit of $R$-current correlators in $AdS$
Supergravity}
\author{J.~Bartels$^{1}$, J.~Kotanski$^{1}$, A.-M.~Mischler$^{1}$, 
V.~Schomerus$^{2}$
\bigskip\\
{\it $^1$~II. Institute Theoretical Physics, Hamburg University, Germany}\\
{\it $^2$~DESY Theory Group, Hamburg, Germany}\\
}

\maketitle

\begin{abstract}
\noindent
Four-point functions of $R$-currents are discussed within Anti-de Sitter 
supergravity. In particular, we compute Witten diagrams with graviton and  
gauge boson exchange in the high energy Regge limit. Assuming validity of 
the AdS/CFT correspondence, our results apply to $R$-current four-point 
functions of ${\cal N}=4$ super Yang-Mills theory at strong coupling.
\end{abstract}

\vspace{-9cm}
\begin{flushright}
{\small DESY--09--118}\\
\end{flushright}
\vspace{8.5cm}

\section{Introduction}

Studies of the Regge limit
for scattering amplitudes go back
to the 1960ies when experiments started to explore the high energy
regime of quantum field theories. In Quantum Chromodynamics (QCD),
the weak coupling limit of the Regge limit turned out to be dominated 
by the BFKL Pomeron \cite{Kuraev:1976ge,Kuraev:1977fs,Balitsky:1978ic}
which represents a bound state of two reggeized gluons.
More general, high energy scattering amplitudes in QCD can be
written in terms of reggeon field theory with  
~reggeized gluons \cite{Gribov:1968fc} as the fundamental degrees of freedom.
The BFKL Pomeron is an intriguing starting point for 
analyzing both the NLO corrections \cite{Fadin:1998py,Ciafaloni:1998gs,
Camici:1997ij} to the BFKL kernel and for generalizing 
the BFKL equation to the more complex BKP states in the $t$ channel 
\cite{Bartels:1980pe,Kwiecinski:1980wb,Jaroszewicz:1980mq}.
In the context of large-$N_c$ limits, the BFKL Pomeron represents the 
leading approximation of the elastic scattering amplitude 
(color singlet exchange). The BKP states have been found to be integrable 
for large $N_c$ \cite{Lipatov:1993yb,Lipatov:1994xy,Faddeev:1994zg}, and 
these links of high energy QCD with integrable models have raised hopes for at 
least partial solutions.

Investigations of the planar Regge limit were mostly performed
perturbatively, i.e.~for small 't Hooft coupling $\lambda$.
Up until 1997, strongly coupled gauge theory has remained largely
inaccessible, at least with analytical tools. The situation has
changed through the discovery of the AdS/CFT correspondence
~\cite{Maldacena:1997re,Witten:1998qj,Gubser:1998bc}. It relates
many interesting superconformal gauge theories to string theories
in Anti-de Sitter backgrounds. The simplest example of such a
correspondence involves ${\cal N}=4$ supersymmetric Yang-Mills
theory in four space-time dimensions. This theory is an attractive
toy model. While being severely constrained by its symmetries, the
leading Regge asymptotics is identical to that of QCD. The dual
description is given by type IIB string theory on $AdS_5 \times
S^5$. Even though the latter is very difficult to solve completely, 
calculations may be performed for large radius $R$ of $AdS_5$ using 
the approximate description through classical supergravity. 
According to the AdS/CFT correspondence, the supergravity limit 
of string theory is dual to gauge theory at strong coupling 
$\lambda \rightarrow \infty$.

In Quantum Chromodynamics, the scattering of electromagnetic
currents provides a reliable environment for studying the 
Pomeron \cite{Bartels:1996ke,Brodsky:1997sd}: at large virtuality of the 
external photons, the QCD coupling constant is small, and the use of 
perturbation theory is justified. ${\cal
N}=4$ super Yang-Mills theory contains close relatives of
electromagnetic current \cite{CaronHuot:2006te}, namely the
$R$-currents which belong to the global $SU_R(4)$ group.
Therefore, is seems natural to further explore the Pomeron, within the 
AdS/CFT correspondence, by
investigating four-point correlators of $R$-currents in ${\cal N}
=4$ super Yang-Mills theory. In the weakly coupled regime, the
relevant correlation functions have been investigated  
\cite{Bartels:2008zy}. Similar to the QCD case, in the high energy limit 
the scattering amplitude has the form of a convolution of the two 
$R$-current impact
factors 
\be
{\cal A}_{P_AP_B}(s,t)=is \int \frac{d^2 k}{(2\pi)^3} \Phi_{P_A}(Q_A^2;\bk,\bq-\bk)  
\frac{1}{\bk^2 (\bq-\bk)^2}
\Phi_{P_B}(Q_B^2;\bk,\bq-\bk)\,,
\label{eq:2gluon}
\ee
where $\bk$, $\bq$ are two-dimensional transverse momentum vectors with $t=- \bq^2$, $Q_A^2$ and 
$Q_B^2$ are the virtualities of the two incoming $R$-currents (for simplicity we take 
the virtualities of the outgoing currents to be identical to the incoming ones).
Helicity is conserved, and $P_A$, $P_B$ denote 
the polarizations of the two incoming external currents (transverse or longitudinal).     
The impact factors $\Phi_A$, $\Phi_B$ for the $R$-currents in ${\cal N}=4$ 
have been calculated explicitly ~\cite{Bartels:2008zy}. 
When including higher order corrections in the leading logarithmic approximation  
the two gluon propagators are replaced by the BFKL Green's function
\ba
{\cal A}_{P_AP_B}(s,t)&=&is \int_{- i\infty}^{i \infty}  \frac{d \omega}{2 \pi i} s^{\omega} 
\int \frac{d^2 k}{(2\pi)^3} \frac{d^2 k'}{(2\pi)^3}
\nonumber \\[2mm] &&
\times 
 \Phi_{P_A}(Q_A^2;\bk,\bq-\bk) 
G(\bk,\bq-\bk;\bk',\bq-\bk';\omega) \Phi_{P_B}(Q_B^2; \bk',\bq-\bk'), 
\label{BFKL}
\ea
and the leading singularity in the $\omega$ plane is located at 
$\omega_0=4  N_c \alpha_s   \ln 2/\pi$.
It may be worthwhile to recall that the high energy behavior (\ref{BFKL}) also allows to address 
the short distance behavior of the operator product of the $R$-currents:
by considering the limit $Q_A^2 \gg Q_B^2$ (like in deep inelastic electron proton scattering), 
the BFKL Green's function can be used to derive, in particular, the anomalous dimension of 
the leading twist two gluon operator expanded around the point $\omega=0$.        

As we recalled in the previous paragraph,
the limit of strong coupling is determined by classical
supergravity. Techniques for the relevant supergravity
computations were developed starting from \cite{Witten:1998qj}. They have
been applied to calculate many two- and three-point correlation
functions
\cite{Freedman:1998tz,D'Hoker:1999ea,Chalmers:1998xr,Bianchi:2003ug}
as well as four-point correlators
\cite{Freedman:1998bj,D'Hoker:1998gd,D'Hoker:1998mz,D'Hoker:1999pj,%
D'Hoker:1999ni,D'Hoker:2000dm,Metsaev:1998hf,Mueck:1998iz,Arutyunov:2000py,%
Arutyunov:2003ae}. Just as in the case of weak coupling, most of
the supergravity amplitudes possess unpleasant divergencies which
must be renormalized
\cite{Bianchi:2001de,Bianchi:2001kw,DeWolfe:2001pq,Freedman:1999gp,%
DeWolfe:2000xi,Bianchi:2000sm,Skenderis:2002wp,Papadimitriou:2004rz,%
Papadimitriou:2004ap}. In evaluating the Regge limit of four-point
amplitudes we actually meet a pleasant surprise. It turns out that
UV-divergent terms do not contribute to the high energy Regge
limit of scattering amplitudes. The main reason is that, in the Regge limit, 
the coordinates of the $R$-currents on the boundaries are well separated 
from each other, thus avoiding the ultraviolet divergencies. 
Therefore, none of the calculations performed below requires 
holographic renormalization.

In this paper we calculate the high energy limit for the
four-point correlation function of $R$-currents at strong
coupling, restricting ourselves to the leading term, i.e. in the supergravity regime. 
We evaluate contributions from graviton and gauge boson exchange in the bulk
and show that the leading Regge asymptotics is determined entirely
by the t-channel exchange of bulk gravitons. All other amplitudes,
including those from s- and u-channel exchange of gravitons and of
gauge bosons, are suppressed by at least one power of the energy.
Let us mention that there exist a variety of other approaches to
high energy scattering at strong
coupling \cite{Polchinski:2000uf,Polchinski:2001tt,%
Polchinski:2002jw,Brower:2006ea,Brower:2007qh},
\cite{Hatta:2007cs,Hatta:2007he,Hatta:2008st} and
\cite{Cornalba:2006xm,Cornalba:2006xk,Cornalba:2007zb,Cornalba:2008qf}.
Our work has a somewhat different take in that is focuses on
$R$-currents. Furthermore, the results are obtained by thoroughly
analyzing the underlying Witten diagrams, without any shortcuts or
additional assumptions.

Let us briefly describe the content of the following sections. We
shall begin with a more expository part in which some of the
relevant background material is presented. In order to define the
high energy limit in Section \ref{sc:FT} all relevant propagators
must be transformed to momentum space. Section 3 is devoted to
a rather detailed calculation of the graviton exchange. The
corresponding scattering amplitude is proportional to the square of
the total energy, $s^2$ \cite{'tHooft:1987rb}. The properties of
the amplitude are investigated in some detail. In particular,  after 
performing an inverse Fourier transform on the transverse momenta 
of the process, we find a simple closed expression in configuration space.
The resulting expression for the Regge limit of the graviton exchange can be
written rather compactly in terms of a scalar propagator of
$AdS_3$. Moreover, we determine the expansion coefficients for 
the amplitude written as a series in the exchanged momentum. We also 
investigate the dependence of the forward scattering amplitude on
the virtualities of the process. In Section \ref{sc:BE} we
calculate the Regge limit of gauge boson exchange 
for general values of the exchanged momentum. As 
expected for the exchange of a vector boson, the amplitude is
proportional to the total energy $s$.

\section{Definitions and ingredients from supergravity}

In this second section we shall formulate our main task we
address and we provide the basic ingredients that are required 
for its completion.  After a short review on the calculation of
$R$-current correlators from supergravity, we list all the necessary
building blocks. These include the bulk-to-bulk propagators for
gravitons and gauge bosons in Anti-de Sitter space.

\subsection{Formulation of the Problem}

We consider ${\cal N}=4$ super Yang-Mills (SYM) theory in
 four dimensional Euclidean space. Let us pick one of its $R$-currents
by $J_{j}$ with $j$ labeling the spacial directions, i.e.\
$j=1,\dots,d=4$. $\vec{x}=(x_1,x_2,x_3,x_4)$ denotes the four 
dimensional Euclidean vector. We are interested in evaluating the Fourier
transform of the four-point correlator,
\be i (2 \pi)^4  \delta(\sum_i\vec p_i) A_{j_1 j_2 j_3 j_4}(\vec
p_i) =\int \left( \prod_{i=1}^4 d^4 x_i \e^{-i \vec p_i \cdot \vec
x_i} \right) \langle J_{j_1}(\vec x_1) J_{j_2}(\vec x_2)
J_{j_3}(\vec x_3) J_{j_4}(\vec x_4) \rangle\,  . \lab{eq:amp} \ee
Due to the conservation of the $R$-current, i.e.\ $\partial_j J^j
=0$, the contraction of the quantity $A$ with one of the four
external momenta vanishes trivially. We can solve these Ward
identities explicitly by projecting the scattering amplitudes,
\be {\cal A}_{\lambda_1\lambda_2;\lambda_3\lambda_4}(|\vec
p_i|;s,t) =\sum_{j_i}
 \epsilon_{j_1}^{(\lambda_1)}(\vec p_1)
 \epsilon_{j_2}^{(\lambda_2)}(\vec p_2)
 \epsilon_{j_3}^{(\lambda_3)}(\vec p_3)^{\ast}
 \epsilon_{j_4}^{(\lambda_4)}(\vec p_4)^{\ast}
{A}_{j_1 j_2 j_3 j_4}(\vec p_i) \,, \quad
\lambda_i=L,\pm \,,
\ee
with appropriate polarization vectors
$\epsilon_{j}^{\lambda_i}(\vec p_i)$ satisfying $p_i^{j}
\epsilon_{j}^{(\lambda_i)}(\vec{p_i}) =0$ along with an
orthonormality condition. A set of polarization vectors with the
required properties is spelled out in appendix A (eq.\ (\ref{eq:pv})). For
any given choice of polarizations, the resulting scattering
amplitude can only depend on the two Mandelstam variables $s$ and
$t$.

The perturbative computation of the full scattering amplitude in
gauge theory and supergravity is possible, though a rather tedious
exercise. In this study we shall only be interested in the Regge
limit of the scattering amplitude ${\cal A}(s,t)$ of the process 
$1+2\to3+4$, i.e. in the limit where 
the total energy is much larger than the momentum transfer and the 
virtualities of the external currents. In Euclidean notation we have 
\be 
s\ =\ -(\vec p_1+\vec p_2)^2\,, -t\ =\ (\vec p_1+\vec p_3)^2\,, 
\ee
and $|\vec p_1|$, $|\vec p_2|$ ($|\vec p_3|$ and $|\vec p_4|$) 
are the virtualities of the incoming (outgoing) currents. In order to 
take the limit we are interested in, namely   
\be
|\vec{p_i}|^2\,, -t \ll s\,,
\lab{eq:rglm}
\ee
we have to go, via Wick rotation, to the Minkowski space. 

Our aim here is to calculate the amplitude (\ref{eq:amp}) in the limit of 
infinite 't Hooft coupling (the weak coupling limit has been addressed 
in \cite{Bartels:2008zy}).  To this
end we make use of the conjectured AdS/CFT correspondence
\cite{Maldacena:1997re} between IIB string theory on $AdS_{d+1}$
space and ${\cal N}=4$ $SU(N_c)$ super Yang-Mills theory. An
efficient calculation can only be performed in the limit of large
$N_c$ (planar limit). At the same time we send the 't Hooft
coupling $\lambda=g^2_{YM} N_c$ to infinity. In this regime, the
full string theory on $AdS_{d+1}$ is well approximated by
classical supergravity.

The AdS/CFT correspondence comes with a prescription to compute
correlation functions in the $d$-dimensional quantum field theory
\cite{Witten:1998qj,Gubser:1998bc}. To be more precise, sources
$\phi_0$ of operators in super Yang-Mills theory correspond to the
boundary values of supergravity fields in $AdS_{d+1}$, i.e
$\phi|_{\partial AdS}\sim\phi_0$. For an n-point function we have
\be \langle J(1) J(2) \ldots J(n) \rangle_{CFT} = \omega_n
\frac{\delta^n}{\delta \phi_0(1) \ldots \delta \phi_0(n)}
\exp(-S_{AdS}[\phi[\phi_0]]) \big|_{\phi_0=0} \,,
\lab{eq:AdSCFTcorr} \ee
where the factor $\omega_n$ comes from the relative normalization
of sources to $\phi_0$ values and the normalization of the action
\cite{Chalmers:1998xr}. On the right hand side, $S_{AdS}$ denotes
a classical supergravity action that is evaluated with fixed
boundary values of $\phi$.

Before we can spell out the supergravity theory we consider, let
us briefly fix some conventions concerning the Anti-de Sitter space
$AdS_{d+1}$. Its Euclidean continuation is parameterized by
$z_0>0$ and $\vec x$ with coordinates $x_i$ enumerated by
the Latin indices $i=1,\ldots,d$. The metric is given by
\be ds^2= \frac1{z_0^2}(d z_0^2+ d \vec x^2)\,, \lab{eq:ds2} \ee
where $d \vec x^2$ can be related to the metric of Minkowski space
by Wick rotation. The boundary of the Anti-de Sitter space is at
$z_0=0$. Our computations will be performed for $d=4$, the case
that is relevant for QCD.

Our supergravity calculations can be truncated consistently to a
theory involving fluctuations of the metric on $AdS_{d+1}$ along
with fluctuations of an $U(1)_R$ gauge field $A_{\mu}$. The latter
is related to the gauge theory $R$-currents through the AdS/CFT
correspondence. The relevant supergravity action reads
\be S=\frac{1}{2 \kappa^2} \int d^{d+1} z \sqrt{g}(-{\cal
R}+\Lambda)+S_m\,, \lab{eq:Sg} \ee
with ${\cal R}$ being the scalar curvature and where the covariant
matter action is
\cite{Freedman:1998tz,Chalmers:1998xr,Cvetic:1999xp,Arutyunov:2000py}
\be S_m= \frac{1}{2\kappa^2} \int d^{d+1} z \sqrt{g}
\left[\frac{1}{4 }   F_{\mu \nu} F^{\mu \nu} + \frac{i k}{24
\sqrt{g}}  \varepsilon_{\mu \nu \rho \sigma \lambda} F_{\mu \nu}
F_{\rho \sigma} A_{\lambda} -A_\mu J^\mu +\ldots \right]\ .
\lab{eq:Sm} \ee
Here $2\kappa^2 = 15 \pi^2 R^3/N^2_c$, $R$ denotes the radius of 
$AdS_5$,  and $F_{\mu \nu}$ is the field strength of the gauge field 
$A$, as usual. Throughout this note, Greek indices refer to the
$(d+1)$-dimensional space, i.e.\ they take values from $0$ to $d$.
Latin subscripts, on the other hand, parameterize directions along
the Euclidean $d$-dimensional boundary of $AdS_{d+1}$. Repeated 
indices are always summed over after they have been lowered. To 
lower indices we use the $d+1$-dimensional metric. The 
coefficient $k$ of the Chern-Simons is an integer.

Evaluation of the four-point correlation function of $R$-currents
using eq.\ (\ref{eq:AdSCFTcorr}) along with eq.\ (\ref{eq:Sg}) is, in
principle, rather straightforward. In practice, we can use a very
convenient and intuitive diagrammatic procedure that was first
proposed by Witten \cite{Witten:1998qj} and then developed further
by many other authors. In our case, the computation of the
relevant Witten diagrams requires only three basic building blocks.
These include the bulk-to-bulk propagators for the graviton and
the gauge $R$-bosons as well as the bulk-to-boundary $R$-boson
propagator. They are connected  by vertices which can be inferred
form  eqs.\ (\ref{eq:Sg}) and (\ref{eq:Sm}). The diagrams that
shall be analyzed below are plotted in Fig.~\ref{fig:gravboson}.
We are only interested in their leading Regge behavior. Since the
regime (\ref{eq:rglm}) is characterized through momenta, it is
necessary to transform the various propagators to momentum space. In
order to make our presentations reasonably self-contained, we
shall list the Fourier transform of the basic building blocks in
the following three subsections.

\lab{sc:FT}
\begin{figure}
\begin{center}
{\epsfysize4.0cm \epsfbox{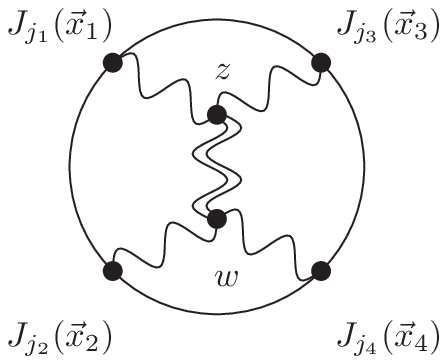}\quad\epsfysize4.0cm \epsfbox{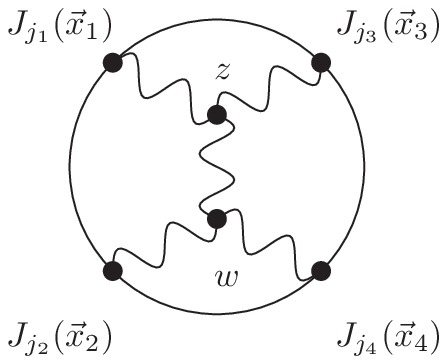}}
\end{center}
\caption{Witten diagrams for the graviton and boson exchange in the $t-$channel, respectively}
\lab{fig:gravboson}
\end{figure}

\subsection{Bulk-to-bulk propagator of the gauge boson}

Let us begin by discussing the Fourier transform of the
bulk-to-bulk gauge boson propagator. According to
\cite{D'Hoker:1999jc}, its coordinate space representation is
given by
\be G_{\mu \nu'}(z,w)= -(\partial_\mu
\partial_{\nu'}u) F(u)+\partial_\mu \partial_{\nu'} S(u)\,,
\ee
where $\partial_{\mu}$ is the derivative with respect to $z_\mu$
and $\partial_{\mu'}$ denotes derivatives with respect to the
components $w_{\mu'}$. The propagation of the physical components
are described by the massive scalar propagator $F$ with mass
parameter $m^2=-(d-1)$,
\be F(u)=\frac{\Gamma\left(\frac{d-1}{2}\right)}{(4
\pi)^{(d+1)/2}}[u(2+u)]^{-(d-1)/2}
=
\frac{\Gamma\left(\frac{d-1}{2}\right)}{(4 \pi)^{(d+1)/2}}
\xi^{d-1}\sum_{k=0}^{\infty}
\frac{\Gamma(k+\ft{d-1}{2})}{\Gamma(\ft{d-1}{2})\Gamma(k+1)}
 \xi^{2k}\,.
\ee
The function $S(u)$, on the other hand, is a gauge artifact. We
shall not need its form below. The so-called chordal distance
variable $u$, finally, is defined by
\be u \equiv  {\frac{(z-w)^2}{2z_0 w_0}}\,, \qquad
\xi=\frac{1}{1+u}=\frac{2 w_0 z_0}{z_0^2 +w_0^2 +(\vec z - \vec
w)^2}\,, \ee
where $(z-w)^2 = \delta_{\mu \nu}(z-w)_{\mu} (z-w)_{\nu}$ is the
so-called flat Euclidean distance. In spelling out the Fourier
transform, we distinguish two different cases according to whether
the second index $\nu'$ is parallel or transverse to the boundary
of $AdS_{d+1}$. In the first case where $\nu'=j$, the result of
the Fourier transform reads
\ba G_{\mu j}(z_0,w_0,\vec q) &=& \int d^4x \e^{i \vec q\cdot \vec
x} G_{\mu j}(z_0,w_0,\vec x) = q_\mu q_{j} \tilde S +
\sum_{k=0}^{\infty} \frac{ 2^{1-d} (z_0 w_0)^{d/2-1}
}{\Gamma(k+d/2) \Gamma(k+1)} \nonumber \\[2mm]
&& \hspace*{-3cm} \times \left[ \delta_{\mu j}
 \left(\frac{z_0^2 w_0^2 \vec q^2}{4\varpi^2_0}\right)^{k+d/4-1/2}
K_{2k+d/2-1}(|\vec q| \varpi_0) -\delta_{\mu 0} \ft{i}{2} q_j w_0
\left(\frac{z_0^2 w_0^2 \vec q^2}{4\varpi_0^2}\right)^{k+d/4-1}
K_{2k+d/2-2}(|\vec q| \varpi_0) \right] \nonumber \,, \ea
where $\vec x=\vec z-\vec w$ and $\tilde S$ is the Fourier transform of $S$. Moreover, we introduced
the function $\varpi_0 = \varpi_0(z_0,w_0) = \sqrt{w_0^2 +
z_0^2}$. Here and below, $K_m(x)$ denotes the modified Bessel
function.

In the second case, when $\nu'=0$, the Fourier transform of the
gauge boson's bulk-to-bulk propagator reads,
\ba G_{\mu 0}(z_0,w_0,\vec q) &=& \int d^4x \e^{i \vec q\cdot \vec
x} G_{\mu 0}(z_0,w_0,\vec x) = q_\mu q_{0} \tilde S +
\sum_{k=0}^{\infty} \frac{2^{1-d} ( z_0 w_0)^{d/2-1}
}{\Gamma(k+d/2) \Gamma(k+1)}
\nonumber \\[2mm]
&& \hspace*{-3cm} \times \left[ \delta_{\mu 0}
\frac{\varpi_0^2}{z_0 w_0}
 \left(\frac{z_0^2 w_0^2 \vec q^2}{4\varpi_0^2}\right)^{k+d/4-1/2}
\!\!\!\!\! K_{2k+d/2-1}(|\vec q| \varpi_0) + \delta_{\mu j}
\ft{i}{2} q_j
 z_0
 \left(\frac{z_0^2 w_0^2 \vec q^2}{4\varpi_0^2}\right)^{k+d/4-1}
 \!\!\!\!\! K_{2k+d/2-2}
 (|\vec q| \varpi_0)
\right. \nonumber \\[2mm] && \left.
\hspace*{3cm} -
 \delta_{\mu 0}
(k+\ft{d}{2}-1)
 \left(\frac{z_0^2 w_0^2 \vec q^2}{4\varpi_0^2}\right)^{k+d/4-1}K_{2k+d/2-2}
 (|\vec q| \varpi_0)
\right] \nonumber  \,. \ea
The symbol $q_0$ in the first term on the right hand side denotes
the derivative  $q_0 \equiv i \partial_0$ with respect to $z_0$.

\subsection{Bulk-to-boundary propagator of the gauge boson}
Following \cite{Witten:1998qj}, let us now consider the
bulk-to-boundary propagator of the gauge boson in Lorentz-like
gauge. It is given by the simple expression
\be G_{\mu j}(z,\vec x)\ =\ N_d \delta_{\mu j} \frac{z_0^{d-2}}{
((\vec z - \vec x)^2+z_0^2)^{d-1}} -
 N_d \delta_{\mu 0}\frac{1}{2(d-2)}
\frac{\partial}{\partial \vec x_j} \frac{z_0^{d-3} }{ ((\vec z -
\vec x)^2+z_0^2)^{d-2}}\,, \ee
where the normalization
$$ N_d\ =\ \ft{(d-2)\Gamma(d)}{2 \pi^{d/2} (d-1)\Gamma(d/2)}$$
is chosen such that the bulk-to-boundary propagator
$G_{jl}(z,\vec x) \to  \delta_{jl} \delta^{(d)}(\vec z-\vec x)$ in the limit where $z_0$ is sent to $z_0 =0$.
The formula we state here is valid for dimensions $d>2$.

Performing a Fourier transform in the spacial coordinate $\vec{x}$
on the boundary, we obtain
\ba G_{\mu j}(z_0,\vec p) &=& \int dz \e^{i \vec p\cdot (\vec
z-\vec x)} G_{\mu j}(z,\vec x) \ = \ N_d \delta_{\mu j } z_0^{d-2}
\frac{2 \pi^{d/2}}{\Gamma(d-1)} \left(\frac{|\vec p|}{2
z_0}\right)^{d/2-1}K_{d/2-1}(z_0|\vec p|) \nonumber \\[1mm]  &&
\hspace*{3cm} -N_d \delta_{\mu 0}  i p_j
 z_0^{d-3}
\frac{ \pi^{d/2}}{\Gamma(d-1)} \left(\frac{|\vec p|}{2
z_0}\right)^{d/2-2}K_{d/2-2}(z_0|\vec p|)\ .  \ea This expression
simplifies considerably when we set $d=4$. In this case, the
normalization $N_4$ takes the form $N_4=2/\pi^2$ and therefore
\ba G_{\mu j}(z_0,\vec p) &=& z_0 \left[ \delta_{\mu j }
 |\vec p| K_{1}(z_0|\vec p|) - i p_j \delta_{\mu 0}
 K_{0}(z_0|\vec p|) \right] \ .  \ea

\subsection{Bulk-to-bulk propagator of the graviton}
Finally, we turn our attention to the bulk-to-bulk propagator of
the graviton. According to \cite{D'Hoker:1999jc} this quantity
reads as
\ba G_{\mu \nu ;\mu' \nu'}& =& (\partial_\mu \partial_{\mu'} u\,
\partial_\nu \partial_{\nu'} u+\partial_\mu \partial_{\nu'}
u\,\partial_\nu \partial_{\mu'} u) \,G(u) +g_{\mu \nu} g_{\mu'
\nu'} \,H(u) + \dots
\lab{eq:graprop} \ea
Here the dots $\dots$ stand for terms of the form $\partial_\rho X$
where $\rho = \mu,\mu',\nu,\nu'$. These turn out not to contribute
to our computation below. Furthermore, $G$ is the massless scalar
propagator
\be G(u)\ =\ 2^d \xi^d C_d \ _2F_1(d/2,(d+1)/2;d/2+1;\xi^2) =\pi
^{-\frac{d+1}{2}} \xi^d \sum_{k=0}^{\infty} \frac{  \Gamma
\left(\frac{d+1}{2}+k\right)}{2 (d+2 k) \Gamma (k+1)} \xi^{2 k}\,,
\ee where
$$ C_d \ =\ \frac{  \Gamma
\left(\frac{d+1}{2}\right)}{(4 \pi)^{(d+1)/2} d }.$$
The function $H$, finally, is given by the following two explicit
formulas
 \ba
H(u)&=&-\frac{2 G(u)}{(d-1) \xi^2} +\frac{2(d-2)}{\xi (d-1)^2} 2
C_d (2 \xi)^{d-1} \ _2F_1((d-1)/2,d/2;d/2+1;\xi^2)
 \\
&=&-\frac{2  }{(d-1)}\frac{\xi^{d-2}}{ \pi ^{\frac{d+1}{2}}}
\sum_{k=0}^{\infty} \frac{  \Gamma \left(\frac{d+1}{2}+k\right)}{2
(d+2 k) \Gamma (k+1)} \xi^{2 k} +\frac{(d-2)}{(d-1)} \frac{
\xi^{d-3} }{  \pi ^{\frac{d+1}{2}}} \sum_{k=0}^{\infty} \frac{
\Gamma \left(\frac{d-1}{2}+k\right)}{2(d+2 k) \Gamma (k+1)} \xi^{2
k}\,. \nonumber \ea %
Before studying the Fourier transform, we evaluate the
quantity
\ba
\partial_\mu \partial_{\mu'} u \ = \
\frac1{z_0 w_0} \delta_{\mu \mu'} +\frac{(z-w)_{\mu} }{ z_0 w_0^2}
\delta_{\mu' 0} +\frac{(w-z)_{\mu'}}{z_0^2 w_0} \delta_{\mu 0}
-\frac{(z-w)^2}{2 z_0^2 w_0^2} \delta_{\mu 0} \delta_{\mu' 0}.
\lab{eq:XXG}
\ea
Let us anticipate that only the first of these terms does actually
contribute to the high energy behavior. With this is in mind we focus
on the relevant part by defining
\ba\lab{eq:gravprop}
G^{(1)}_{\mu \nu; \mu' \nu'}(z_0,w_0,\vec x) \equiv \left( \
\frac1{(z_0 w_0)^2} \delta_{\mu \mu'}  \delta_{\nu \nu'} +
\frac1{(z_0 w_0)^2} \delta_{\mu \nu'} \delta_{\nu \mu'}\right)
G(u).
\ea
Its Fourier transform is easily computed and reads
\ba
G^{(1)}_{\mu \nu; \mu' \nu'}(z_0,w_0,\vec q) &=& \int d^4x \e^{i
\vec q\cdot \vec x} G^{(1)}_{\mu \nu; \mu' \nu'}(z_0,w_0,\vec x) =
(\delta_{\mu \mu'}\delta_{\nu \nu'}+\delta_{\mu \nu'}\delta_{\nu \mu'}
)
\lab{eq:p4grav}
\\[2mm] &&
\times\sum_{k=0}^{\infty}
\frac{
(z_0 w_0)^{2k+d-2}
}{ \Gamma(k+d/2+1) \Gamma (k+1)}
 \left(\frac{|\vec q|^2}{4\varpi_0^2}\right)^{k+d/4}K_{2k+d/2}
 (\varpi_0|\vec q|)\,,
\nonumber \ea where $\vec x= \vec z- \vec w$, as before.
Subleading terms can be obtained in a similar way, but they will not be 
needed in the present context. We finally mention that, for large $|q|$, the Fourier transform 
of the function $G(u)$ goes as $1/|q|^2$ \ \cite{Brower:2007qh}. 

\section{Graviton exchange in the high energy limit}

The aim of this section is to compute and analyze high energy limit of the 
scattering amplitude ${\cal A}(s,t)$ introduced in the previous section. 
Since we are interested in the $AdS_5$ case, we shall set $d=4$ from now on.
We switch to Minkowski metric $g = \mbox{diag}(+,-,-,-)$.
However, for simplicity we continue using our previous notation:
i.e. $\vec p_j$ now stands for the four vector $(p_{j;4},p_{j;1},p_{j;2},p_{j;3})$ with 
$|\vec p_j|^2= -p_{j;4}^2 + p_{j;1}^2 +p_{j;2}^2 + p_{j;3}^2 = -p_j^2$,   
Latin indices continue to run from $1$ to $4$ (where the fourth component 
denotes Minkowski 'time'), and Greek indices run between $0$ and $4$.  

The first subsection contains the main results on the high energy
limit. In the second subsection we make an attempt to
re-interpret the result as coming from a correlation function in
$AdS_3$. Finally, we investigate in some detail the properties of
the forward scattering at infinite 't Hooft coupling.

\subsection{The graviton exchange}
Let us begin by computing the contribution to the four-point function
of $R$-currents that is obtained from the exchange of a single
graviton in the t-channel (left figure in \ref{fig:gravboson}). According to the rules
of the AdS/CFT correspondence, this quantity is given by\footnote{The
correlation functions and amplitudes are calculated up to multiplicative constants, which can be easily restored from the action (\ref{eq:Sg}).}
\ba 
I^{\rm GR}&=& \frac1{4} \int \frac{d^{4}z d z_0}{z_0} \int
\frac{d^{4}w d w_0}{w_0} T_{(13) \mu \nu}(z) G_{\mu \nu; \mu'
\nu'}(z,w) T_{(24) \mu' \nu'}(w)\,. \lab{eq:Ixgrav} 
\ea 
We shall
refer to $T_{(ij)}$ as the stress-energy tensor. It is determined
through the bulk-to-boundary propagators of the gauge boson and it
contains the coupling between gauge bosons and gravitons.
\ba T_{(13)\mu \nu}&=&
z_0^2
\partial_{[\mu} G_{\lambda] j_1}(z,\vec x_1)
\partial_{[\nu} G_{\lambda] j_3}(z,\vec x_3)
+
z_0^2
\partial_{[\nu} G_{\lambda] j_1}(z,\vec x_1)
\partial_{[\mu} G_{\lambda] j_3}(z,\vec x_3)
\nonumber \\[2mm] && \hspace*{3cm}
-
\frac1{2}
z_0^2
\delta_{\mu \nu}
\partial_{[\alpha} G_{\beta] j_1}(z,\vec x_1)
\partial_{[\alpha} G_{\beta] j_3}(z,\vec x_3)
\ \ . \lab{eq:Tmn} \ea Note that $T$ satisfies the
four-dimensional Ward identity by construction. We shall see below
that only the first two terms of $T_{(13) \mu \nu}$ contribute to
the high energy behavior of the amplitude. Performing the Fourier
transform of the expression (\ref{eq:Ixgrav}) we obtain
\ba 
\tilde I^{\rm GR}(\vec{p}_i)&=& 
\int
\prod_i d^4x_i \e^{- i \sum_j \vec p_j \cdot \vec x_j} 
I^{\rm GR}(\vec{x}_i)\ = \  (2 \pi)^{4}\,
\delta^{(4)}(\sum_i\vec p_i)\ 
\\[2mm] & & \hspace*{-2cm} \times \
\frac1{4} \int \frac{d z_0}{z_0} \int \frac{d w_0}{w_0}  \tilde T_{(13) \mu
\nu}(z_0,\vec p_1,\vec p_3) \tilde T_{(24) \mu' \nu'}(w_0,\vec
p_2,\vec p_4) G_{\mu \nu; \mu' \nu'}(z_0,w_0;\vec p_1+\vec p_3)
\,, \lab{eq:Ipgrav} 
\ea
 where
\be T_{(13)\mu\nu}= \frac{1}{(2\pi)^8} \int d^4 p_1 d^4 p_3 \e^{i
\vec p_1  \cdot (\vec x_1 - \vec z)} \e^{i \vec p_3 \cdot (\vec
x_3 - \vec z)} \tilde T_{(13) \mu \nu}\,. \ee Before we analyze
which terms give the leading contributions to the high energy
behavior, we recall that we still need to contract our amplitude
with the appropriate polarization vectors
\be {\cal\tilde I}^{\rm GR}_{\lambda_1,\lambda_2
\lambda_3,\lambda_4 }  \ = \ {\sum}_{j_i}
 \epsilon_{j_1}^{(\lambda_1)}(\vec p_1)
 \epsilon_{j_2}^{(\lambda_2)}(\vec p_2)
 \epsilon_{j_3}^{(\lambda_3)}(\vec p_3)^{\ast}
 \epsilon_{j_4}^{(\lambda_4)}(\vec p_4)^{\ast}
({\tilde I}^{\rm GR})_{j_1 j_2 j_3 j_4}\,. \ee Since the
amplitude (\ref{eq:Ipgrav}) satisfies the Ward identities
associated with the conservation of $R$-currents, we are allowed to
shift the polarization vectors (\ref{eq:pv}) by the momenta of the
corresponding particles. If we allow for this additional freedom,
the polarization vectors may be brought into the form displayed in
eq.\ (\ref{eq:spv}) of the Appendix. We note that contractions of
these shifted polarization vectors with any tensor cannot give
additional powers of the energy $s$. Hence, we can determine the
dominant terms of the scattering amplitude before we actually
switch to the polarization basis. After these remarks let us look
back at the form of the stress-energy tensor $T$. Each of the
three terms contains two derivatives which are replaced by momenta
after Fourier transformation. These are combined with two more
momentum components from the second stress-energy tensor.
Therefore, we can at most obtain terms which are of the order
$s^2$. But this requires that momentum components of $\vec{p}_1$
and $\vec{p}_3$ are contracted with momentum components
$\vec{p}_2$ and $\vec{p}_4$. Terms in which $\vec{p}_1$ is
contracted with $\vec{p}_3$, on the other hand, are clearly
subleading. This implies that we can drop the term in the second
line of eq.\ (\ref{eq:Tmn}) and it explains why we had previously
introduced the quantity $G^{(1)}$ in our discussion of the
bulk-to-bulk propagator for the graviton. In fact, $G^{(1)}$
contains all terms of the propagator that can contribute to the
leading high energy behavior. We can summarize the results of our
discussion through the following two formulas
\be 
\label{eq:nonsensehelicity}
\tilde G^{(1)}_{ij;\,i'j'}(z_0,w_0,\vec{q})  \approx 
\frac{4 }{s^2 (z_0w_0)^2} \left({p_2}_i {p_1}_{i'} {p_2}_{j}
{p_1}_{j'} +{p_2}_i
 {p_1}_{j'} {p_2}_j
{p_1}_{i'}\right)\tilde G\, ,
\ee
\ba
{\tilde T}_{(13) \mu \nu}(z_0,\vec p_1,\vec p_3) & \approx&
z_0^4
(
 \delta_{\mu k_1}\delta_{\nu k_3}
+ \delta_{\nu k_1}\delta_{\mu k_3} )
 {p_1}_{k_1}
 {p_3}_{k_3} \ \times
 \nonumber \\[2mm] && \hspace*{-2cm}
\times\  \left[
  {p_1}_{j_1}
 {p_3}_{j_3}
 K_{0}(z_0|\vec p_1|)
 K_{0}(z_0|\vec p_3|)
-
\delta_{j_1 j_3 }
 |\vec p_1|
 |\vec p_3|
K_{1}(z_0|\vec p_1|) K_{1}(z_0|\vec p_3|) \right]\, \lab{eq:T13}
\label{eq:impact1}
\ea
for the high energy limit of the graviton bulk-to-bulk propagator
and the stress-energy tensor, respectively. Here and throughout
the rest of the paper, $\approx$ means equality up to terms that
are subleading in the high energy limit. Note that the graviton
propagator has exactly the form that is expected in the Regge
limit: for the exchange of a spin one gauge boson it is well-know that the leading 
high energy behavior comes from a particular $t$-channel helicity state. If $j$ ($j'$) 
denote the upper (lower) Lorentz indices of the $t$-channel exchange propagator and $p_1$ 
($p_2$) the large momenta at the upper (lower) vertex, this dominant  
helicity state contributes through the tensor 
\be
\frac{2 {p_2}_{j} {p_1}_{j'}}{s}.
\ee    
In eq.\ (\ref{eq:nonsensehelicity}) we see that the leading behavior of the graviton exchange 
can be interpreted as the (symmetrized) tensor product of two spin one bosons.    
Furthermore, as we will demonstrate below the first and the second term within the
square bracket in the last line of eq.\ (\ref{eq:impact1}) correspond to longitudinal and
transverse polarization of the $R$-boson, respectively.

If we now substitute the two expressions (\ref{eq:T13}) and
(\ref{eq:p4grav}) back into the amplitude (\ref{eq:Ipgrav}) and
use that $(\vec p_1 \cdot \vec p_2) (\vec p_3 \cdot \vec p_4)
\approx s^2/4$, we arrive at the high energy limit of the
graviton exchange:
\ba \tilde I^{\rm GR}_{\rm Regge} \ =(2 \pi)^4 
\delta^{(4)}(\sum_i \vec p_i)\, \frac{s^2}{2} \, \int d z_0 \int d w_0
\Phi_{j_1j_3}(p_1,p_3;z_0)\Sigma(|\vec p_1+\vec p_3|, z_0,w_0)
\Phi_{j_2j_4}(p_2,p_4;w_0)\,
\lab{eq:ILO1}
\ea
where
\ba \Sigma(|\vec p_1+\vec p_3|, z_0,w_0) & = & \sum_{k=0}^{\infty} \frac{z_0^{2 k+5}
w_0^{2 k+5}
 }{\Gamma(k+1) \Gamma(k+3)}
 \left(\frac{|\vec p_1+\vec p_3|^2}{4\varpi_0^2}\right)^{k+1}
K_{2k+2}(|\vec p_1+\vec p_3| \varpi_0)\,,
\lab{eq:sumLO}
\\[2mm]
 \Phi_{j_1j_3}(\vec p_1, \vec p_3;z_0)& = & \sum_{m=0,1}
 \tilde W^{m}_{j_1 j_3}(\vec p_1,\vec p_3)
 K_{m}(z_0|\vec p_1|) K_{m}(z_0|\vec p_3|)
\, ,   \nonumber 
\ea
while 
\be 
\tilde W^{m_1}_{j_1 j_3}(\vec p_1,\vec p_3) =(\delta_{j_1 j_3}
|\vec p_1| |\vec p_3|
\delta_{m_1,1}-{p_1}_{j_1}{p_3}_{j_3}\delta_{m_1,0})\,.
\lab{eq:Wjj}
\ee
This formula has reminiscent of eq.\ (\ref{eq:2gluon})
where $\Phi_{j_aj_b}$ plays a role of an impact factor
while $\Sigma(|\vec p_1+\vec p_3|, z_0,w_0)$
is the analog of a propagator.

It is convenient to switch to the helicity basis:
by contraction with the polarization vectors 
from eqs.\ (\ref{eq:spv}) and (\ref{eq:eT}) and making use of 
the orthonormality of the transverse polarizations we obtain
\ba 
\lab{eq:Wll}
\tilde {\cal W}^{m_1}_{\lambda_1\lambda_3} (\vec p_1,\vec p_3)
&=& \sum_{j_1,j_3} \epsilon^{(\lambda_1) }_{j_1}(\vec p_1)
\epsilon^{(\lambda_3)}_{j_3}(\vec p_3)^{\ast} \tilde W^{m_1}_{j_1
j_3}(\vec p_1,\vec p_3) \nonumber \\[2mm] &\approx& |\vec p_1|
|\vec p_3| (\delta_{m_1,1}
\delta_{\lambda_1,h}\delta_{\lambda_3,h}+\delta_{m_1,0}
\delta_{\lambda_1,L}\delta_{\lambda_3,L}) \, , \ea
i.e. the first term with $m_1=1$ only contributes to the transverse polarizations 
$h=\pm$, whereas $m_1=0$ belongs to the longitudinal polarization. 
From eq.\ (\ref{eq:Wll}) we learn that helicity is conserved: $\lambda_1 = \lambda_3$.
As we also see in eq.\ (\ref{eq:Wll}),
the quantity $\tilde W_{\lambda_1, \lambda_3}$ in the helicity basis 
only depends upon the virtualities of the external currents. 
Consequently, also the ``impact factor'' in the helicity basis
\be
\Phi_{\lambda_1 \lambda_3}(|\vec p_1|, |\vec p_3|;z_0) =  \sum_{m=0,1}
 \tilde {\cal W}^{m}_{\lambda_1 \lambda_3}(\vec p_1,\vec p_3)
 K_{m}(z_0|\vec p_1|) K_{m}(z_0|\vec p_3|)\,,
\label{eq:impacthelicity}
\ee
only depends on the virtualities.

With these expressions our scattering amplitude finally reads:
\be
\label{eq:gravexch}
{\cal A}_{\lambda_1 \lambda_2 \lambda_3 \lambda_4}^{\rm GR}(s,t)= \frac{s^2}{2} \int dz_0 dw_0 
\Phi_{\lambda_1 \lambda_3}(|\vec p_1|, |\vec p_3|;z_0) \,
 \Sigma(|\vec p_1+\vec p_3|, z_0,w_0) \, 
\Phi_{\lambda_2 \lambda_4}(|\vec p_2|, |\vec p_4|;w_0)\,.
\ee  

The amplitude (\ref{eq:gravexch}) is proportional to $s^2$. To be complete, one 
also has to consider the exchange of the graviton in the $s$- and $u$-
channels. In these cases the last term of the stress-energy tensor
(\ref{eq:Tmn}) is also important. Counting powers of momenta in
the stress-energy tensor one might at first expect to get
additional contributions of order $s^2$. But, unlike in the
$t$-channel exchange that we have discussed at length, the $s$-
and $u$-channel exchanges of the graviton are suppressed by an
additional factor $s^{-1}$ that comes in through the graviton
propagator itself. Therefore, there is no need to analyze such
contributions to the amplitude any further.

Equation  (\ref{eq:gravexch}) should be compared with the weak coupling result (\ref{eq:2gluon}).
Again we have the structure of two impact factors $\Phi$ which depend upon the virtualities of the 
external currents, connected by an exchange propagator $\Sigma$ and 
convoluted by a two-dimensional integration. The power of $s$ reflects the spin of the exchanged graviton.
On the gauge theory side, in the weak coupling limit, the amplitude is 
given by the exchange of two gluons, and higher order corrections in $g^2$ 
replace the two gluon exchange by the BFKL Green's function, modifying the 
power of $s$ from $1$ to $1+\omega_0$. On the string side 
it has been argued that, due to the reggeization of the graviton, the power behavior $s^2$ of the graviton 
exchange will be modified to $s^{2-\Delta}$ where $\Delta = {\cal O}(1/\sqrt{\lambda})$. 
However, in order to compute $\Delta$, one has to go beyond the 
supergravity approximation used in this paper.

\subsection{Going back to configuration space}

In the previous subsection we determined the leading Regge
asymptotics for the Fourier transform of the four-point correlator of
$R$-currents. The result was expressed in terms of the four momenta
$\vec p_i$ and it involved an infinite summation in the
construction of the kernel functions $\Sigma$. In principle, one
might attempt to perform the inverse Fourier transform and to
re-phrase our result as an expression for the four-point correlator
of $R$-currents. But the answer turns out to be rather
complicated.

We will therefore take a different route and perform an inverse
Fourier transform exclusively in the transverse momenta. To make
this more precise, we fix a particular
frame: working in the Minkowski metric $g_{\mu\nu}=\mbox{diag}(1,-1,-1,-1)$ 
we take 
the large momenta along the 3-axis. 
We introduce the light-like reference vectors $p_A =(p,0,0,p)$ and 
$p_B=(p,0,0,-p)$ with $s= 4 p^2$, and the transverse momentum vectors 
$p_{i;\perp}=(0,p_{i;1},p_{i;2},0)$ ($i=1,...,4$), $q_{\perp}=(0,q_1,q_2,0)$ 
with $p_{i;\perp}^2 = -\bp_i^2$, $q^2= -\bq^2$. Throughout this subsection, 
we adopt the four vector notation $p_{i}=(p_{i;4},p_{i;1},p_{i;2},p_{i;3})$, 
and we take the momenta $p_3$ and $p_4$ to be outgoing. Momenta in bold face 
refer to the 2-dimensional transverse space. For large $s$ we find 
\ba
p_1&=&\left( 1+\frac{Q_2^2 +p_{2;\perp}^2
    -\ft1{2}(p_{1;\perp}+p_{2;\perp})^2}{s} \right) p_A - 
      \frac{Q_1^2 + p_{1;\perp}^2}{s} p_B + p_{1;\perp}\,,\nonumber\\[2mm]
p_2&=&-\frac{Q_2^2 +p_{2;\perp}^2}{s} p_A + 
      \left(1 + \frac{Q_1^2 +p_{1;\perp}^2-\ft1{2}(p_{1;\perp}+p_{2;\perp})^2}{s} \right) p_B + p_{2;\perp}\,,\nonumber\\[2mm]
p_3&=&\left( 1+\frac{Q_4^2 + p_{4;\perp}^2
-\ft1{2}(p_{3;\perp}+p_{4;\perp})^2}{s} \right) p_A 
      -\frac{Q_3^2 +p_{3;\perp}^2}{s} p_B + p_{3;\perp}\,,\nonumber\\[2mm]
p_4&=&-\frac{Q_4^2 + p_{4;\perp}^2}{s} p_A + 
      \left(1 + \frac{Q_3^2 +p_{3;\perp}^2
-\ft1{2}(p_{3;\perp}+p_{4;\perp})^2}{s} \right) p_B + p_{4;\perp}
\nonumber \ea
with $ p_{1;\perp} -  p_{3;\perp} =  p_{4;\perp} - p_{2;\perp}$.
Here we introduced the virtualities $Q^2_i = |\vec p_i|^2$, and for the momentum transfer $t$ we have 
\ba
t = 
 (p_{1;\perp} - p_{3;\perp})^2 
+\frac{(Q_2^2 - Q_4^2+p_{2;\perp}^2-p_{4;\perp}^2)(Q_3^2 - Q_1^2+p_{3;\perp}^2-p_{1;\perp}^2)}{s} 
+\ldots \approx\  - \bq^2\ .
\label{eq:tdefinition}
\ea
We now compute the two-dimensional Fourier transform with respect to $q$:
\ba 
{\cal I}^{\rm GR}_{\rm T,Regge}&=&
\frac{1}{(2 \pi)^8}
\int \, \prod_i d^2 p_i \, \prod_i e^{i(\bx_1 \cdot \bp_1 + \bx_2 \cdot  \bp_2 
-\bx_3 \cdot  \bp_3 - \bx_4 \cdot  \bp_4) }
\nonumber \\
&&
\times
 (2\pi)^2 
\delta^{(2)}(\bp_1 + \bp_2 - \bp_3 - \bp_4)\,{\cal A}^{GR}_{\lambda_1 \lambda_2 \lambda_3 \lambda_4}(s,t)\,.
\lab{eq:FT2d} 
\ea
Here, the two dimensional transverse vectors $\bx_i$ define the 
positions in transverse space of the scattering $R$-currents. 
In particular, the difference $\bb = \frac{1}{2} (\bx_1+ \bx_3) - 
\frac{1}{2} (\bx_2+ \bx_4)$ is the impact parameter. 
The subscript $T$ indicates that the Fourier transform 
has been performed in the transverse space only.
In calculating the two-dimensional Fourier transform we recall
that $s$ and $|\vec p_i|$ are kept fixed so that the integration
over transverse momenta only effects the arguments of the delta
function and the variable $t = - |\vec p_1 + \vec p_3|^2$ in 
eq.\ (\ref{eq:tdefinition}). 
The result is
\ba {\cal I}^{\rm GR}_{\rm T,Regge}&\approx&
\delta^{(2)}(\bx_1-\bx_3)
\delta^{(2)}(\bx_2-\bx_4)
\lab{eq:amtg}
\\[2mm]&&
\times
\frac{s^2}{2} \, \int d z_0
z_0^2 \int d w_0 w_0^2
\Phi_{\lambda_1 \lambda_3}(|\vec p_1|, |\vec p_3|;z_0) 
G_{\Delta=3,d=2}(\bar u)
\,\Phi_{\lambda_2 \lambda_4}(|\vec p_2|, |\vec p_4|;w_0).
\ea 
The variable $\bar u = (1-\bar \xi)/\bar \xi$ is related to a
3-dimensional analogue of our variable $\xi$ through
$$ \bar
\xi \ =\ \ft{2 z_0 w_0 }{z_0^2+w_0^2+\bb^2}\ . $$
The function $G$ that appears in formula (\ref{eq:amtg}) is
obtained as a special choice of the scalar Green's function in
$AdS_{d+1}$
\cite{Fronsdal:1974ew,Burgess:1984ti,Inami:1985wu,Burges:1985qq}
\be G_{\Delta,d}(u)\ =\ 2^\Delta \frac{\Gamma(\Delta)
\Gamma(\Delta-\ft{d}{2}+\ft{1}{2})}{(4 \pi)^{(d+1)/2} \Gamma(2
\Delta-d+1)} \xi^{\Delta} \
_2F_1(\ft{\Delta}{2},\ft{\Delta+1}{2};\Delta-\ft{d}{2}+1;\xi^2)\,.
\lab{eq:GuDd} \ee
Our parameters $\Delta = 3$ and $d=2$ are associated with a scalar
of mass $m^2=\Delta(\Delta-d)=3$ in $AdS_3$. Standard properties
of hypergeometric functions allow to simplify the expression for
this particular propagator to read
\be G_{\Delta=3,d=2}(\bar u)= \frac{2-\bar \xi^2-2 \sqrt{1-\bar
\xi^2}}{4 \pi  \bar \xi \sqrt{1-\bar \xi^2}}\,. \lab{eq:G32} \ee

Note that the dependence of the scattering
amplitude on the 2-dimensional positions $\bx_i$ is rather
simple, in that the transverse positions of the incoming currents are 
conserved: $\bx_1 = \bx_3$ and $\bx_2=\bx_4$.
In our computation, the delta functions appear while taking the high energy
limit. Our formula (\ref{eq:amtg}) can be used to study the 
b-dependence of the high energy scattering amplitude. 
For large impact parameter, we expand in inverse powers of $b^2$, and  
the leading terms has the exponent $\delta = - 6$. 
In the regime in which $b^2$ is small, the scattering amplitude is 
proportional to $\log b^2$.
Numerical curves of the amplitude (\ref{eq:amtg}) are plotted in
Fig.~\ref{fig:xXamp}.

\begin{figure}
\begin{center}
{
\psfrag{'xXTT.dat'}{\small $X_{TT}(\bb)$}
\psfrag{'xXLT.dat'}{\small $X_{LT}(\bb)$}
\psfrag{'xXLL.dat'}{\small $X_{LL}(\bb)$}
\psfrag{x34}{$\mbf{b}$}
\epsfysize6.5cm \epsfbox{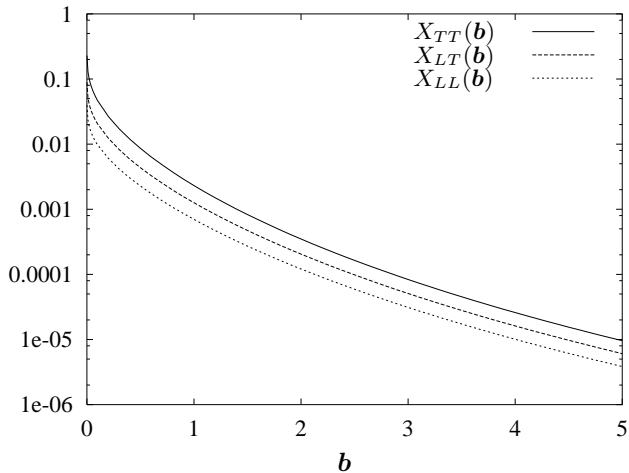}}
\end{center}
\caption{Integrals of the amplitude (\ref{eq:amtg}) plotted as a function of $\bb$ and fixed $Q_i=1$. Different lines correspond to different polarizations of $R$-bosons taken from the Bessel function subscripts, namely $1 \to T$ and $0 \to L$. For $\bb \to \infty$ the amplitudes $X_{TT}$, $X_{TL}=X_{LT}$, $X_{TT}$ vanish as $\ft{32}{25 \pi}\bb^{-6}$, $\ft{64}{75 \pi}\bb^{-6}$, $\ft{128}{225 \pi}\bb^{-6}$, respectively.}
\lab{fig:xXamp}
\end{figure}

\subsection{Investigation of the forward amplitude}

Before we spell out and study the formula for the amplitude at
$t=0$ it is useful to re-expand the function $\Sigma$ that was
introduced in eq.\ (\ref{eq:sumLO}) in the vicinity of vanishing
$-t = |\vec q|^2 = |\vec p_1 + \vec p_3|^2$. This may be achieved
with the help of eq.\ (\ref{eq:bessel}). The resulting series
\ba
\Sigma(|\vec q|,z_0,w_0) &=& \sum_{m=0}^{\infty} |\vec q|^{2m} U_{m}
\,, \lab{eq:Tser} 
\ea
is often helpful to read off properties of the Regge limit
(\ref{eq:rglm}). Its expansion coefficients $U_m$ are spelled out
in Appendix \ref{sc:sigma} in terms of hypergeometric functions.
Only the first summand $U_0$ is needed for the forward amplitude:
\ba 
{\cal A}^{\rm GR}_{\lambda_1 \lambda_2 \lambda_3 \lambda_4} (s,t=0) &=& 
s^2 \int_0^{\infty} d z_0 z_0^3
\Phi_{\lambda_1 \lambda_3}(|\vec p_1|, |\vec p_3|;z_0) \,
\nonumber \\
&&
\times
\int_0^{\infty} d w_0 w_0^3
\Phi_{\lambda_2 \lambda_4}(|\vec p_2|, |\vec p_4|;w_0) 
\ft1{2}G_{\Delta=2,d=0}(\hat u)\,, 
\lab{eq:LOfr} 
\ea
where $\hat u=(z_0-w_0)^2/2 z_0 w_0$, and the propagator
\be
G_{\Delta=2,d=0}(\hat u)=\frac{1}{4 w_0^2 z_0^2}
(\theta(w_0-z_0)z_0^4+\theta(z_0-w_0)w_0^4), 
\label{eq:t=0grav}
\ee
is again a special case of the scalar propagator in $AdS_{d+1}$
that we had introduced previously in eq.\ (\ref{eq:GuDd}).

\begin{figure}
\begin{center}
{
\psfrag{'XTT.dat'}{\small $\tilde X_{TT}(r)$}
\psfrag{'XLT.dat'}{\small $\tilde X_{LT}(r)$}
\psfrag{'XTL.dat'}{\small $\tilde X_{TL}(r)$}
\psfrag{'XLL.dat'}{\small $\tilde X_{LL}(r)$}
\psfrag{rp}{$r$}
\epsfysize6.5cm \epsfbox{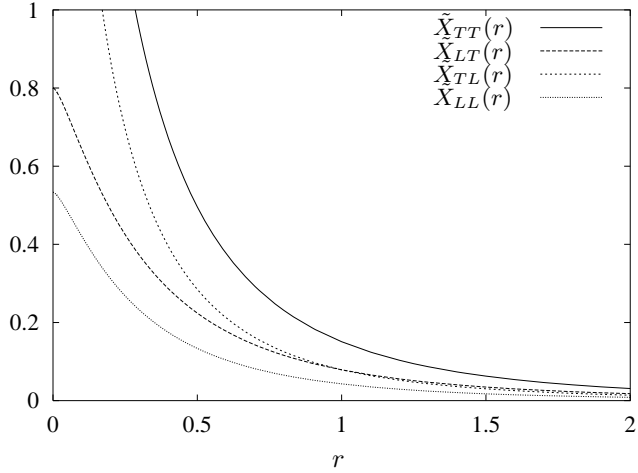}}
\end{center}
\caption{The amplitude functions defined by Eq.~(\ref{eq:XPP}) plotted as a function of $r=Q_A/Q_B$.}
\lab{fig:Xamp}
\end{figure}

In the following evaluation of the forward amplitude we restrict 
ourselves to two independent virtualities denoted by $Q_A^2=|\vec p_1|^2=|\vec
p_3|^2$ and $Q_B^2=|\vec p_2|^2=|\vec p_4|^2$. First we are interested in the
behavior of the high energy amplitude as a function of the ratio $r
= Q_A/Q_B$. We start from eq.\ (\ref{eq:LOfr}). From eq.\ (\ref{eq:Wll}) we know 
that helicity is conserved, i.e. we have the four possibilities that the 
helicities $\lambda_1=\lambda_3=P_A$ and $\lambda_2=\lambda_4=P_B$ are transverse or longitudinal.
Let us denote these cases by $P_A P_B = TT$, $LT$, $TL$, and $LL$, resp. Correspondingly we write:  
\be
{\cal A}^{\rm GR}_{P_A P_B}(s,t=0) = \frac{s^2}{8 Q_B^4} \tilde X_{P_AP_B}\,,
\lab{eq:DIS}
\ee
where   
\be
\tilde X_{P_AP_B}(r) =  r^2 \int_0^{\infty} d z_B  d w_B
(\theta(w_B-z_B)z_B^5 w_B+\theta(z_B-w_B) w_B^5 z_B)
K_{m_A}^2(z_B r) K_{m_B}^2(w_B) \,,
\lab{eq:XPP} 
\ee
and $P_A,P_B$ stand for the two polarizations $T,L$. The index of the 
Bessel function is given by $m_A =1$ when $P_A = T$ and by $m_A = 0$ 
for $P_A = L$. Our integration variables $z_B$  and $w_B$ are related 
to $z_0$ and $w_0$ through $z_B = z_0 Q_B$ and $w_B = w_0 Q_B$.

The functions (\ref{eq:XPP}) are plotted in
Fig.~\ref{fig:Xamp} as a function of the ratio $r=Q_A/Q_B$, 
for all choices of polarizations.  When $r\to 0$, 
the integrals $\tilde X_{LL}$ and $\tilde X_{LT}$ go to the finite values
$$\tilde X_{LL}(0) = \ft{8}{15} \ \ \ \ \mbox{ and }\ \ \
 \tilde X_{LT}(0)= \ft{8}{10}\ \ . $$
The other two integrals $\tilde X_{TL}$ and $\tilde X_{TT}$, on
the other hand, diverge logarithmically,  i.e.\
$$ \tilde X_{TL} (r) \sim \ft{8}{15} \ln
r^{-2}\ \ \ \mbox{ and } \ \ \ \tilde X_{TT}(r) \sim \ft{8}{10}\ln
r^{-2}\ \  $$
for small $r \sim 0$. 
The behavior of the amplitudes at large $r$ can easily be 
obtained
using the symmetry with respect to the exchange of $p_A$ and
$p_B$. Our definition of the quantities $\tilde X$ respects the
exchange symmetry only up to an overall factor $r^4$, i.e.\
\be 
\tilde X_{P_A P_B}(1/r)\ =\ r^4 \tilde X_{P_BP_A}(r)\  .  
\ee
From this we can immediately read off the asymptotic behavior of
$\tilde X$ at large values of $r$. The functions $\tilde X_{LL}$
and $\tilde X_{TL}$ vanish like $r^{-4}$ while $\tilde X_{LT}$ and
$\tilde X_{TT}$ behave as $r^{-4} \ln r^2$. 

The region of large $r=Q_A/Q_B$ corresponds to the limit of 'deep inelastic scattering'
of the upper $R$-current on the lower $R$-current, i.e. the upper 
current plays the role of the photon, the lower one that of the target. 
Since in the Witten diagram approximation our scattering amplitude has no imaginary part, 
we cannot compute a cross section; nevertheless, it may be instructive to 
analyze the power behavior in $r$. Introducing the scaling variable 
$x_{bj}\approx Q_A^2/s$,
and combining, in eq.\ (\ref{eq:DIS}), the power $r^{-4}$ of 
$\tilde X_{P_A P_B}(r)$  with the factor $s^2$ in front of the integral, we 
find for large $r$ the leading behavior 
\be
{\cal A}^{GR}_{P_AP_B} \sim \left(\frac{1}{x_{bj}}\right)^2\,,   
\ee
for $(P_A P_B)$= LL and TL, whereas 
\be
{\cal A}^{GR}_{P_AP_B} \sim \left(\frac{1}{x_{bj}}\right)^2 \ln r^2\,,   
\ee
for the cases LT and TT.
As we have indicated after eq.\ (\ref{eq:2gluon}), the limit 
$r^2 = Q_A^2/Q_B^2 \to \infty$ 
is connected with the short distance behavior of the product of the $R$-currents 
$J(\vec x_1)$ and $J(\vec x_3)$. 

It may be of interest to compare this behavior with the large $r=Q_A/Q_B$ behavior of the weak 
coupling limit of the forward scattering amplitude in ${\cal N}=4$ SYM in eq.\ (\ref{eq:2gluon}). 
For large $r=Q_A/Q_B$, 
the maximal power of logarithms in $r$ comes from the region of strongly ordered 
transverse momenta: $Q_A^2 \gg {\bk}^2 \gg Q_B^2$. In addition, the impact factors 
$\Phi_A$ and $\Phi_B$, depending upon the polarization of the external 
currents, may contain logarithms of the type $\ln Q_A^2/\bk^2$ and 
$\ln \bk^2/ Q_B^2$. From ~\cite{Bartels:1996ke,Brodsky:1997sd} we take:
\ba
\Phi_{A,P_A} (Q_A^2,\bk^2)&\sim& 
\left\{ \begin{array}{cc} \frac{Q_A^2}{\bk^2} \ln \frac{Q_A^2}{\bk^2}&\, : \, P_A=T\\
                          \frac{\bk^2}{Q_A^2}&\, : \,  P_A=L 
\end{array} 
\right.  
\nonumber\\
\Phi_{B,P_B} (Q_B^2,\bk^2)&\sim& 
\left\{ \begin{array}{ll} \ln \frac{\bk^2}{Q_B^2}&\, : \, P_B=T\\
                           \mbox{const} &\, : \, P_B=L \end{array} \right. \,.
\ea 
This leads to 
\ba
{\cal A}_{P_AP_B} \sim \left\{ \begin{array}{ll} c_{TT} \cdot \ln^3 r^2 &\, : \,  (P_AP_B)=(TT) \\
                                          c_{TL} \cdot \ln^2 r^2 &\, : \,  (P_AP_B)=(TL) \\
                                          c_{LT} \cdot \ln^2 r^2 &\, : \,  (P_AP_B)=(LT) \\
                                          c_{LL} \cdot \ln r^2 &\, : \,  (P_AP_B)=(LL) 
                                          \end{array} \right.\,, 
\label{Astructure}
\ea
and the constants are composed of the coefficient functions and 
anomalous dimensions. Returning to the graviton exchange amplitude, 
it is suggestive to associate the $\ln r^2$ modification also with 
anomalous dimensions: it would be 
interesting to perform a systematic operator analysis of the strong coupling limit.     

In the last part of this section we would like to determine the region in the integration
over $z_0$ and $w_0$ from which the four amplitudes $\tilde X$
receive their dominant contributions.
Let us consider the integrands of the functions $\tilde X_{PP'}$ 
which we now write as
\be 
\tilde X_{PP'}\ =\int_0^{\infty} d z_B \int_0^{\infty} d w_B \tilde J_{PP'}\,. 
\ee
As one can see from eq.\ (\ref{eq:XPP}), the integrand is finite at 
$z_B=0$ and $w_B=0$, and due to the Bessel functions it falls off exponentially at infinity. One therefore 
expects, for $Q_A \sim Q_B$, the main contribution to the integrals to come 
from the region where $z_B$ and $w_B$ are of order unity. 
There are two kinds of maxima: first, there is a 'ridge' along the diagonal line 
$z_B = w_B$, resulting from $\theta$-functions in eq.\ (\ref{eq:XPP}). 
Second, there are local maxima away from the diagonal. 
Maxima along the ridge are found from the condition 
\be
\partial_{x} \tilde J_{P_AP_B}(z_B=w_B=x)|_{x=x_{P_AP_B}} = 0,
\ee
which has the form:
\be 0= 2(m_A+m_B+3) K_{m_A}(r x)K_{m_B}(x) -2 v  K_{m_A}(r x)
K_{m_B+1}(x) -2 r x   K_{m_A+1}(r x) K_{m_B}(x)\,, 
\lab{eq:pos}
\ee
whereas the other extrema satisfy the conditions
\ba
 \partial_{w_B} \tilde J_{P_AP_B}&=&
r^2 w_B^4  z_B K_{m_A}(r z_B)^2 K_{m_B}(w_B) \left((2 m_B+5)
K_{m_B}(w_B)-2 w_B   K_{m_B+1}(w_B)\right)
=0\,, \nonumber \\[2mm]
\partial_{z_B} \tilde J_{P_AP_B}&=&
r^2 w_B^5 K_{m_A}(r z_B)    K_{m_B}(w_B)^2 \left((2 m_A+1)
   K_{m_A}(r z_B)-2 r z_B K_{m_A+1}(r z_B)\right)
=0\,. \nonumber 
\ea
We remind that $m_A$ ($m_B$) refer to the helicities of the upper (lower) 
$R$-currents, and $m=1$ ($m=0$) belongs to transverse (longitudinal) polarizations.
In the following we describe the results of our analysis which are illustrated 
in  Fig.~\ref{fig:max}. 
\begin{figure}
\begin{center}
{ \psfrag{LL}{$LL$} \psfrag{LT}{$LT$} \psfrag{TL}{$TL$}
\psfrag{TT}{$TT$} \psfrag{yLL}{\!\!\!\!\!\!\!\!\!\small
$y_B^{LL}=y_B^{TL}$} \psfrag{LTTT}{$TT$, $LT$} \psfrag{pa0}{$Q_A
\to 0$} \psfrag{painf}{\rotatebox{45}{$\infty \leftarrow Q_A \to
0$}} \psfrag{vTT}{$v_{TT}$} \psfrag{vTL}{$v_{TL}$}
\psfrag{yBLL}{$y_{B,LL}$} \psfrag{yBLT}{$y_{B,LT}$}
\psfrag{vA}{$z_B=z_0\, Q_B$} \psfrag{vB}{$w_B=w_0\, Q_B$}
\epsfysize7.0cm \epsfbox{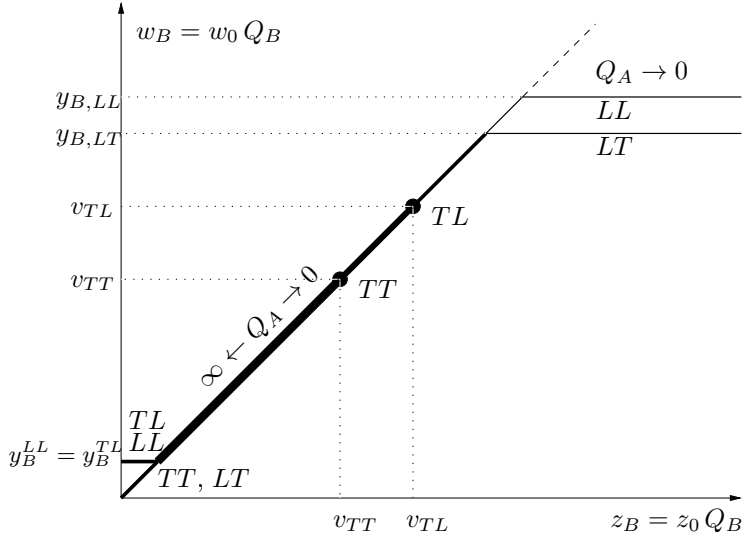}}
\end{center}
\caption{The position of the integrand maximum in the $(z_0,w_0)-$space  as a function of $Q_A$ and with $Q_B=$ fixed
for different $R$-boson polarizations. The black circles denote 
points corresponding to $Q_A \to 0$  for polarizations $TT$ and 
$TL$.} 
\lab{fig:max}
\end{figure}

Let us begin with the special case in which the two
virtualities $Q_A$ and $Q_B$ are equal, i.e.\ $r=1$. 
The maximum of the integrand lies on the diagonal line line at 
$z_B=w_B=x_{P_AP_B}$: in Fig.~\ref{fig:max} these points lie to the left of the
points labeled by 'TT'. The numerical values of $v_{TT} \approx 1.3316$ and 
$v_{TL} \approx 1.5527$, 
as expected, are close to unity. This implies that, in the graviton exchange Witten diagram in
Fig.\ref{fig:gravboson}, the coordinates $z_0$ and $w_0$ of the 
interaction vertices are both of the order $1/Q_B$.

Next, let us fix $Q_B$ and consider the amplitude as a function of
$r=Q_A/Q_B$. 
When $r$ decreases, the maxima on the ridge at $z_B=w_B=x_{P_AP_B}$ move 
upwards along the diagonal. For the polarizations TT and TL, they reach, for  
$r=0$, their final values at the points marked by 'TT' and 'TL', resp.
For the other two cases, LL and LT, the maxima continue to move along the 
diagonal ray until they reach, for some value $r^{(l)}$, their turning values 
denoted by $y_{B}^{LL}$ and $y_{B}^{LT}$. After that they continue along the 
horizontal lines to the right of these turning points. For small $r$, their 
$z_B$ coordinate grows as $1/r$. 
 
Next we consider the limit $r\to \infty$. Beginning again at $r=1$ on the 
diagonal line and increasing $r$, the maxima $z_B=w_B=x_{P_AP_B}$ now move in the opposite 
direction, towards the origin. For the cases TT and LT they stay on the 
diagonal until they reach the origin, and for large $r$ the values 
$x_{P_AP_B}$ decrease as 
$1/r$. For the polarizations LL and TL, there are again turning points at 
$y_{B,LL}=y_{B,LT}$, and beyond these points the maxima move on the horizontal line  
towards the $w_B$ axis.\footnote{These conclusions for the large-$r$ behavior 
can be obtained from the observation 
that the integrands $I_{P_AP_B}$ are invariant when 
replacing $r$ by $1/r$, i.e. $\tilde I_{P_AP_B}(1/r) \leftrightarrow r^2 
\tilde I_{P_BP_A}(r)$. 
This symmetry can then be used to evaluate the amplitudes for large
values of the ratio $r$.} 

Let us visualize these results in the graviton exchange diagram in Fig.\ref{fig:gravboson}. 
When $Q_A$ becomes large (at fixed $Q_B$), we have found that $z_0 \sim 1/Q_A$ becomes small, i.e. the vertex 
where the exchanged graviton couples to the gauge boson moves close to the boundary, for both polarizations
$P_A=$T and $P_A=$L. For the lower vertex we see a difference between transverse and longitudinal 
polarization of the target current: in the former case, $w_0 \sim 1/Q_A$ becomes small,
in the latter case $w_0$ remains constant of the order $1/Q_B$.   
  
\section{Gauge boson exchange in the high energy limit}
\lab{sc:BE}

In the following pages we shall briefly analyze contributions from
gauge boson exchange in the bulk. In analogy to the case of graviton
exchange, the leading contribution arises from the $t$-channel
exchange of the gauge boson. Not surprisingly, this leading term
is linear in $s$ and, hence, subleading when compared to the
graviton exchange.

We restrict ourselves to the abelian part of the $SU(4)$ group.   
The coupling of the external gauge boson to the exchanged bulk
boson then proceeds only through the Chern-Simons interaction term
\be \int d^dz d z_0 \varepsilon_{\mu \nu \rho \sigma \lambda}
\partial_{\mu} A_{\nu}(z)
 \partial_{\rho}
A_{\sigma}(z) A_{\lambda}(z)\,. \ee
Using once more the standard rules of the AdS/CFT correspondence,
we can determine the amplitude for boson exchange in $t-$channel
to be of the form
\ba I^{\rm CS}&=&
 \int d^4z d z_0  \int d^4w d w_0
 \varepsilon_{\mu_1 \mu_2 \mu_3 \mu_4 \mu_5}
 \varepsilon_{\nu_1 \nu_2 \nu_3 \nu_4 \nu_5}
\partial_{[\mu_1} G_{\mu_2] j_{1}}(z, \vec x_{1})
\partial_{[\mu_3} G_{\mu_4] j_{3}}(z,\vec x_{3})
 \nonumber \\[2mm] && \hspace*{4cm}\times
G_{\mu_5 \nu_5}(z,w)
\partial_{[\nu_1} G_{\nu_2] j_{2}}(w, \vec x_{2})
\partial_{[\nu_3} G_{\nu_4] j_{4}}(w,\vec x_{4}) \,,
\lab{eq:CSt} \ea
where the partial derivatives $\partial$ acting on $z$ and $w$,
respectively. There are similar amplitudes including derivatives
on the bulk-to-bulk propagator. But all these contributions to the
full $t$-channel exchange turn out to be identical, as a result of the
Bianchi identity. As in the case of the graviton exchange, we
perform the Fourier transform, i.e.~
\ba I^{\rm CS}=
\frac1{(2 \pi)^{16}}
\int d^4p_1 d^4p_2 d^4p_3 d^4p_4 \e^{i \vec p_{1}\cdot  \vec
x_{1}} \e^{i \vec p_{2}\cdot  \vec x_{2}} \e^{i \vec p_{3}\cdot
\vec x_{3}} \e^{i \vec p_{4}\cdot  \vec x_{4}} \tilde I^{\rm
CS}\,.
\ea
After tedious calculations one finds that, in the high energy limit,
the main contribution comes from Eq.~(\ref{eq:CSt}) with both
nonzero $\mu_5=m \neq 0$ and $\nu_5=n \neq 0$, i.e.~
\ba I^{\rm CS}_{\rm Regge}&=& 4 \int d^4z d z_0   \int d^4w d w_0
 \varepsilon_{0 \mu_2 \mu_3 \mu_4 m}
 \varepsilon_{0 \nu_2 \nu_3 \nu_4 n}
\partial_{[0} G_{\mu_2] j_{1}}(z, \vec x_{1})
\partial_{[\mu_3} G_{\mu_4] j_{3}}(z,\vec x_{3})
\nonumber \\[2mm] && \hspace*{1.5cm} \times\
G_{m n}(z,w)
\partial_{[0} G_{\nu_2] j_{2}}(w, \vec x_{2})
\partial_{[\nu_3} G_{\nu_4] j_{4}}(w,\vec x_{4})
\nonumber \\[3mm] &&
+\begin{pmatrix}
\vec x_1 \leftrightarrow \vec x_3 \\
 j_1 \leftrightarrow j_3 \\
\end{pmatrix}
+
\begin{pmatrix}

\vec x_2 \leftrightarrow \vec x_4 \\
 j_2 \leftrightarrow j_4 \\
\end{pmatrix}
+
\begin{pmatrix}
\vec x_1 \leftrightarrow \vec x_3; \vec x_2 \leftrightarrow \vec
x_4\\
 j_1 \leftrightarrow j_3;  j_2 \leftrightarrow j_4 \\
\end{pmatrix}
\,. \ea
The Fourier transform of this expression is given by 
\ba \tilde
I^{\rm CS}_{\rm Regge}&=& (2 \pi)^{4}
\delta^{(4)}(\vec p_{1}+\vec p_{2} +\vec p_{3}+\vec p_{4}) \tilde
W^{\rm CS}_{j_1 j_3 j_2 j_4} 
\nonumber \\[2mm] && \times 
\left\{  \int  d z_0 z_0^{3} \int  d w_0 w_0^3
|\vec p_3| K_{1}(z_0|\vec p_3|)  K_{0}(z_0|\vec p_1|)
\right. \nonumber \\[2mm]&&
\times \left. \sum_{k=0}^{\infty}
\frac{ 
K_{2k+1}(|\vec q| \varpi_0)
 }{\Gamma(k+2) \Gamma(k+1)}
 \left(\frac{z_0^2 w_0^2|\vec q|^2}{4\varpi_0^2}\right)^{k+1/2}
\ |\vec p_4| K_{1}(w_0|\vec p_4|)  K_{0}(w_0|\vec p_2|)  
\right\}
\nonumber \\[2mm] &&\hspace*{1cm}
+
\begin{pmatrix}
\vec p_1 \leftrightarrow \vec p_3 \\
 j_1 \leftrightarrow j_3 \\
\end{pmatrix}
+
\begin{pmatrix}
\vec p_2 \leftrightarrow \vec p_4 \\
 j_2 \leftrightarrow j_4 \\
\end{pmatrix}
+
\begin{pmatrix}
\vec p_1 \leftrightarrow \vec p_3; \vec p_2 \leftrightarrow \vec p_4\\
 j_1 \leftrightarrow j_3; j_2 \leftrightarrow j_4 \\
\end{pmatrix}
\,, \lab{eq:CSI} \ea
where the polarization tensor $\tilde W$ takes following form
\ba
 \tilde W^{\rm CS}_{j_1 j_3 j_2 j_4}
&\approx&
t s {p_1}_{j_1} {p_2}_{j_2}     \delta_{j_4 j_3 }
-
s   |\vec p_1|^2 |\vec p_2|^2
(\delta_{j_2 j_3 }  \delta_{j_1 j_4 } -\delta_{j_1 j_2 } \delta_{j_3 j_4 })
+s {p_1}_{j_1} {p_2}_{j_2}
q_{j_3}
q_{j_4}
\nonumber \\[2mm] &&
-s |\vec p_1|^2 {p_2}_{j_2}
(\delta_{j_1 j_4} q_{j_3}-\delta_{j_3 j_4} q_{j_1})
-s |\vec p_2|^2 {p_1}_{j_1} (\delta_{j_3 j_4}
q_{j_2}-\delta_{j_2 j_3} q_{j_4}) \lab{eq:WCSIII} \, 
\ea with
$\vec q=\vec p_1+\vec p_3$, as before. Going to the polarization
base (\ref{eq:spv}) the tensor $\tilde W^{\rm CS}$ gets replaced
by
\ba  \tilde {\cal W}^{\rm CS}_{\lambda_1 \lambda_3,\lambda_2
\lambda_4}(\vec p_i)&=& \sum_{j_1,j_2,j_3,j_4}
\epsilon^{(\lambda_1)}_{j_1}(\vec p_1)
\epsilon^{(\lambda_2)}_{j_2}(\vec p_2)
\epsilon^{(\lambda_3)}_{j_3}(\vec p_3)^{\ast}
\epsilon^{(\lambda_4)}_{j_4}(\vec p_4)^{\ast}
 \tilde W^{\rm CS}_{j_1 j_3 j_2 j_4}(\vec p_i)
 \\[2mm]
&\approx&
-s
|\vec p_1|^2  |\vec p_2|^2
(
(\vec \epsilon^{(\lambda_1)}_{T} \cdot \vec \epsilon^{(\lambda_4)\ast}_{T})
(\vec \epsilon^{(\lambda_2)}_{T} \cdot \vec \epsilon^{(\lambda_3)\ast}_{T})
-
(\vec \epsilon^{(\lambda_3)\ast}_{T} \cdot \vec \epsilon^{(\lambda_4)\ast}_{T})
(\vec \epsilon^{(\lambda_1)}_{T} \cdot \vec \epsilon^{(\lambda_2)}_{T})
)
\nonumber \\[2mm]
&=& -s |\vec p_1|^2  |\vec p_2|^2 
(2 \delta^{\lambda_1, \lambda_2}-1) \delta^{\lambda_1
\lambda_3} \delta^{\lambda_2 \lambda_4}\,, \nonumber \ea
where $\lambda_i$ are  possible transverse polarizations. Putting this
back into our expression (\ref{eq:WCSIII}) we have now expressed the
Regge limit in terms of the kinematical invariants
$s,t,|\vec{p}_i|$. The whole amplitude is proportional to $s$, as
we anticipated at the beginning of this section. Therefore, it is
subleading when compared to the contribution from $t$-channel
graviton exchange. We also note that only the transverse
polarization contributes to the Regge limit of the $t$-channel
exchange of gauge bosons. Furthermore, the helicity is conserved
in the high energy limit.

As in the case of graviton scattering
(\lab{eq:FT2d}) we can perform
two-dimensional Fourier transform in the transverse momenta
to get
\ba 
{\cal I}_{\rm T,Regge}^{\rm CS}&=&  
 \delta^{(2)}(\bx_1-\bx_3)
 \delta^{(2)}(\bx_2-\bx_4)
{\cal W}^{\rm CS}_{\lambda_1 \lambda_3 \lambda_2 \lambda_4}
\nonumber \\[2mm] &&\hspace*{-1cm}  \times  \int
d z_0 z_0^{2} \int  d w_0 w_0^2
|\vec p_3| K_{1}(z_0|\vec p_3|)  K_{0}(z_0|\vec p_1|)
\ G_{\Delta=2,d=2}(\tilde u) 
\ |\vec p_4| K_{1}(w_0|\vec p_4|) K_{0}(w_0|\vec p_2|)  
\nonumber \\[2mm] && +
\begin{pmatrix}
\vec p_1 \leftrightarrow \vec p_3 \\
 \lambda_1 \leftrightarrow \lambda_3 \\
\end{pmatrix}
+
\begin{pmatrix}
\vec p_2 \leftrightarrow \vec p_4 \\
 \lambda_2 \leftrightarrow \lambda_4 \\
\end{pmatrix}
+
\begin{pmatrix}
\vec p_1 \leftrightarrow \vec p_3; p_2 \leftrightarrow \vec p_4\\
 \lambda_1 \leftrightarrow \lambda_3; \lambda_2 \leftrightarrow \lambda_4 \\
\end{pmatrix}\,
\lab{eq:CSIIIRegge} \ea
with $W^{\rm CS}_{j_1 j_3 j_2 j_4}\sim s$ defined similarly to
$W^{m_1}_{j_1 j_3}$ and the scalar propagator $G_{\Delta,d}$ that
was spelled out in eq.~(\ref{eq:GuDd}). Finally, we would like to
provide an expression for the forward limit, i.e.\ when $|\vec p_1
+\vec p_3|=|\vec q| \to 0 $. In this limit, the terms in the curly
brackets of eq.\ (\ref{eq:CSI}) give
\ba \{\ldots\}&=&
-\ft{1}{2}
\int  d z_0 \int  d w_0
|\vec p_3| K_{1}(z_0|\vec p_3|) K_{0}(z_0|\vec p_1|)
\ (\theta(w_0-z_0) z_0^4 w_0^2 + \theta(z_0-w_0) z_0^2 w_0^4)
\nonumber \\[2mm] && \hspace*{2cm}\times
\ |\vec p_4|  K_{1}(w_0|\vec p_4|) K_{0}(w_0|\vec p_2|) 
\,. 
\ea
Thus, for $\vec p_1=-\vec p_3$ $\vec p_2=-\vec p_4$, we are left
with
\ba 
\tilde I^{\rm CS}_{\rm forward} &=&   (2 \pi)^{4}
\delta^{(4)}(\vec p_{1}+\vec p_{2} +\vec p_{3}+\vec p_{4}) (8\  \vec
p_1 \cdot \vec p_2)   \left(
\delta_{j_1 j_2 }    \delta_{j_3 j_4 }-\delta_{j_2 j_3 }
\delta_{j_1 j_4 }   \right) \nonumber \\[2mm] && \hspace*{-2cm}
\times    \int  d z_0 \int  d w_0 z_0^3 w_0^3
\  |\vec p_1|^3 K_{0}(z_0|\vec p_1|) K_{1}(z_0|\vec p_1|) 
 \ G_{\Delta=1,d=0}(\hat u) 
\  |\vec p_2|^3 K_{0}(w_0|\vec p_2|)
K_{1}(w_0|\vec p_2|)
\lab{eq:CSforward}
\,. 
\ea 
The scalar propagator $G$ in the second line is defined
through eq.\ (\ref{eq:GuDd}). It assumes the following form
\be G_{\Delta=1,d=0}(\hat u)=\frac{1}{2 z_0 w_0} (\theta(w_0-z_0)
z_0^2 +\theta(z_0-w_0) w_0^2)\,. \ee

We conclude that the exchange of the gauge bosons
in the $t-$channel  can give contributions to the scattering amplitude
at most proportional to $s$, and it is of the same order as  
subleading terms of the graviton exchange.

\section{Summary}

The aim of this work was to calculate in the Regge limit  the scattering amplitude
of $R$-currents for ${\cal N}=4$ SYM theory at strong coupling.
We make use of the AdS/CFT conjecture which maps this amplitude
on four-point correlation functions of $R$-bosons
in supergravity which live in the Anti-de Sitter space. We have computed Witten diagrams of
graviton and boson exchanges in the $t-$channel. Similar to gauge theories, the exchange 
of highest spin dominates, i.e. the leading contribution comes from the (real-valued) 
Witten diagram with graviton $t$-channel exchange. This confirms the duality between 
the two-gluon exchange in ${\cal N}=4$ SYM and the graviton exchange in the strong coupling region.   
On the gauge theory side, the sum of leading energy logarithms replaces the two gluon exchange 
by the BFKL Pomeron. On the strong coupling side, it has been argued that the 
$s^2$ behavior of the graviton exchange is replaced by $s^{2-\Delta}$ where 
$\Delta = {\cal O}(1/\sqrt{\lambda})$: however, the computation of this correction $\Delta$  
cannot been done in the supergravity approximation used in the present paper.
For spin-1 boson exchange, which is due to the Chern-Simons
vertices, we found the expected high behavior proportional to $s$; helicity is conserved,
and amplitude with longitudinally polarized bosons are subleading for
large $s$. 

For the graviton exchange we have found that, 
in the transverse $(2+1)-$dimensional configuration space, 
one can re-formulate the scattering amplitude as
the exchange of an effective field, build from a scalar field with dimension
$\Delta=3$. 

We have also analyzed how the graviton exchange amplitude
depends upon the virtualities of the external currents.
For large $r^2=Q_A^2/Q_B^2$, the power behavior is the same as the leading twist  
behavior on the weak coupling side, and the appearance of a logarithm, $\ln r^2$, 
hints at the presence of an anomalous dimension in a short distance expansion. 
A systematic study should be done in a separate paper.    

The virtualities of the external currents determine the distance of the vertices away from the 
boundary. In particular, we have analyzed in some detail the limit $r\to \infty$, i.e. 
the analogue of 'deep inelastic scattering' where the virtuality of the upper $R$-current is much larger than 
the lower one: in this case, the distance of the upper 'impact factor' from the boundary is 
small, of the order $1/r$. A similar result has been found in \cite{Hatta:2007cs,Hatta:2007he,Hatta:2008st}.

We view our results as a first step towards a systematic calculation of the  
Regge limit of the scattering amplitude at strong coupling. In particular, in order to 
obtain a nonzero imaginary part, one has to go beyond the tree approximation of Witten diagrams.
In \cite{Polchinski:2000uf,Polchinski:2001tt,Polchinski:2002jw,Brower:2006ea}
a semiclassical approximation of string theory has been proposed. Another possible line of 
investigation might follow the classical paper of Amati et. al ~\cite{Amati:1987wq} in which string theory 
in flat space has been investigated.

\section*{Acknowledgments}
We are grateful for discussions with  A.~H.~Mueller, L.~Motyka and 
in particular with I. Papadimitriou. This work was supported by the 
grant of SFB 676, Particles, Strings and the Early Universe:
``the Structure of Matter and Space-Time'' and the grant of
the Foundation for Polish
Science.

\appendix
\section{Euclidean polarization vectors}
The polarization vectors  $\vec \epsilon^{(i)}_{k}(\vec p_j)$
should satisfy
\be
\vec p_j \cdot \vec \epsilon^{(i)}(\vec p_j)=0\,,
\qquad
\vec \epsilon^{(i_1)}(\vec p)
\cdot
\vec \epsilon^{(i_2)}(\vec p)^{\ast}=(-)^{L_{i_1,i_2}}\delta^{i_1 i_2}\,,
\lab{eq:polcon}
\ee
where $L_{i_1,i_2}=1$  when $i_j$  is longitudinal $(L)$ and $0$ when $i_j$  is  transverse $(h=\pm)$.
The {\em star} denotes complex conjugation.
One of the possible solutions reads as
\ba
&\vec \epsilon^{(L)}(\vec p_1)=
\frac{1}{|\vec p_1|}(\vec p_1
+\frac{2 |\vec p_1|^2}{s} \vec p_2 )\,,
\qquad  & \vec \epsilon^{(h)}(\vec p_1)=\vec \epsilon_T^{(h)}
+ \frac{2 \vec p_1 \cdot \vec \epsilon_T^{(h)}}{s} \vec p_2\,,
\nonumber\\
&\vec \epsilon^{(L)}(\vec p_2)=
\frac{1}{|\vec p_2|}(\vec p_2
+\frac{2 |\vec p_2|^2}{s} \vec p_1 )\,,
\qquad  & \vec \epsilon^{(h)}(\vec p_2)=\vec \epsilon_T^{(h)}
+ \frac{2 \vec p_2 \cdot \vec \epsilon_T^{(h)}}{s} \vec p_1\,,
\nonumber\\
&\vec \epsilon^{(L)}(\vec p_3)
=
\frac{1}{|\vec p_3|}(-\vec p_1
-\frac{2 |\vec p_3|^2}{s} \vec p_2 +\vec q )\,,
\qquad  & \vec \epsilon^{(h)}(\vec p_3)
=\vec \epsilon_T^{(h)}
+\frac{2 (\vec p_1-\vec q) \cdot \vec \epsilon_T^{(h)}}{s} \vec p_2\,,
\nonumber\\
&\vec \epsilon^{(L)}(\vec p_4)=
\frac{1}{|\vec p_4|}(-\vec p_2
-\frac{2 |\vec p_4|^2}{s} \vec p_1-\vec q )\,,
\qquad  & \vec \epsilon^{(h)}(\vec p_4)
=\vec \epsilon_T^{(h)}
+\frac{2 (\vec p_2+\vec q) \cdot \vec \epsilon_T^{(h)}}{s} \vec p_1\,,
\lab{eq:pv}
\ea
plus subleading terms which do not contribute.
We denote the transfered momenta $\vec q= \vec p_1+\vec p_3$ while
\be
\epsilon_T^{(\pm)}=\frac{1}{\sqrt{2}}(0,1,\pm i,0)\,.
\lab{eq:eT}
\ee
In Eq.~(\ref{eq:pv}) we have shown only the leading terms in $s$.
Using the Ward identity one can shift the polarization vectors to
\ba
&\vec \epsilon^{(L)}(\vec p_1)= \frac{2 |\vec p_1|}{s} \vec p_2\,,
\qquad  & \vec \epsilon^{(h)}(\vec p_1)=\vec \epsilon_T^{(h)}
+ \frac{2 \vec p_1 \cdot \vec \epsilon_T^{(h)}}{s} \vec p_2\,,
\nonumber\\
&\vec \epsilon^{(L)}(\vec p_2)= \frac{2 |\vec p_2|}{s} \vec p_1
\qquad  & \vec \epsilon^{(h)}(\vec p_2)=\vec \epsilon_T^{(h)}
+ \frac{2 \vec p_2 \cdot \vec \epsilon_T^{(h)}}{s} \vec p_1\,,
\nonumber\\
&\vec \epsilon^{(L)}(\vec p_3)= -\frac{2 |\vec p_3|}{s} \vec p_2\,,
\qquad  & \vec \epsilon^{(h)}(\vec p_3)
=\vec \epsilon_T^{(h)}
+\frac{2 (\vec p_1-\vec q) \cdot \vec \epsilon_T^{(h)}}{s} \vec p_2\,,
\nonumber\\
&\vec \epsilon^{(L)}(\vec p_4)= -\frac{2 |\vec p_4|}{s} \vec p_1 \,,
\qquad  & \vec \epsilon^{(h)}(\vec p_4)
=\vec \epsilon_T^{(h)}
+\frac{2 (\vec p_2+\vec q) \cdot \vec \epsilon_T^{(h)}}{s} \vec p_1\,.
\lab{eq:spv}
\ea
The resulting vectors (\ref{eq:spv}) do not satisfy
eq.\ (\ref{eq:polcon}). For amplitudes which satisfies the Ward identity the contraction with polarization vectors of eq.\ (\ref{eq:spv})  gives
the same results as with eq.\ (\ref{eq:pv}). Moreover, since their contraction with other tensors can not
produce additional powers of $s$, they are much simpler in use.
Changing the sign of the metric and performing the Wick rotation one can get
the similar polarization vectors in the Minkowski space.
\section{Momentum space}
\lab{sc:bessel}

Following \cite{Chalmers:1998xr},  Feynman integrals $I$ in 
momentum space can be calculated using the Schwinger representation, i.e.~
\ba
I_m(x_0,\vec q_1)
&=&
\int d^d \vec x_1 \e^{i \vec q_1 \vec x_1} \frac{1}{(x_0^2+\vec x_1^2)^{m+1}}
=
\int d^d \vec x_1 \e^{i \vec q_1 \vec x_1}
\left(
\frac1{\Gamma(m+1)}
\int_0^\infty
d \tau
\tau^m
\e^{-\tau(\vec x_1^2+x_0^2)}
\right)
\nonumber \\
&=&
\frac{\pi^{d/2} }{\Gamma(m+1)}
\int_0^\infty
d \tau
\tau^{m-d/2}
\e^{-\tau x_0^2}
\e^{ - \frac{\vec q_1^2}{4 \tau}  }
\nonumber \\
&=&
\frac{
2 \pi^{d/2}
}{\Gamma(m+1)}
\left(\frac{|\vec q_1|}{2 x_0}
\right)^{m+1-d/2}
K_{m+1-d/2}(x_0 |\vec q_1|)\,.
\ea
For integer positive $\nu$ the modified Bessel function reads as
\ba
K_\nu(u)&=&1/2 (u/2)^{-\nu}
\sum_{k=0}^{\nu-1}
\frac{(\nu-k-1)!}{k!}(-u^2/4)^k
+(-1)^{\nu+1}\ln(u/2)
(u/2)^{\nu}
\sum_{k=0}^{\infty}
\frac{(u^2/4)^k}{k! (\nu+k)!}
\nonumber \\
&&+
(-1)^{\nu}
1/2
(u/2)^{\nu}
\sum_{k=0}^{\infty}
(\psi(k+1)+\psi(\nu+k+1))
\frac{(u^2/4)^k}{k!(\nu+k)!}\,.
\lab{eq:bessel}
\ea

\section{The resummed contribution to the graviton propagator}
\lab{sc:sigma}

The sum defined in  eq.\ (\ref{eq:sumLO})
with $q=|\vec p_1+\vec p_3|$ and $\varpi_0=\sqrt{w_0^2+z_0^2}$
has a following form
\ba
\Sigma(z_0,w_0)&=&
\sum_{m=0}^{\infty}
q^{2m} U_m
=
\sum_{k=0}^{\infty}
\frac{ z_0^{2 k+5} w_0^{2 k+5}  }{\Gamma(k+1) \Gamma(k+3)}
 \left(\frac{q^2}{4\varpi_0^2}\right)^{k+1}K_{2k+2}(q \varpi_0)
\nonumber\\
&=&
\frac{z_0^5 w_0^5}{2\varpi_0^{4}}
\sum_{m=0}^{\infty}
\frac{(-1)^{2m} q^{4m} \varpi_0^{4m}}{2^{4m}\Gamma(2m+1)}
\sum_{k=m}^{\infty}
\frac{ \Gamma(2k+2-2m)  }{ \Gamma(k+1) \Gamma(k+3)}
\frac{ z_0^{2 k} w_0^{2 k}  }{\varpi_0^{4k}}
 \nonumber \\ &&
-
\frac{q^2  z_0^5 w_0^5}{2^3\varpi_0^2}
\sum_{m=0}^{\infty}
\frac{(-1)^{2m} q^{4m} \varpi_0^{4m}}{2^{4m}\Gamma(2m+2)}
\sum_{k=m}^{\infty}
\frac{ \Gamma(2k+1-2m)  }{ \Gamma(k+1) \Gamma(k+3)}
\frac{ z_0^{2 k} w_0^{2 k}  }{\varpi_0^{4k}}
 \nonumber \\ &&
-
\ln(q\varpi_0/2)
\frac{z_0^5 w_0^5 q^4}{2^{4}}
\sum_{m=0}^{\infty}
\frac{(q/2)^{4m}}{ \Gamma(3+2 m)}
\sum_{k=0}^{m}
\frac{\varpi_0^{4m-4k}}{ \Gamma(2 m-2k+1)}
\frac{(-1)^{2k}   z_0^{2 k} w_0^{2 k}
  }{\Gamma(k+1) \Gamma(k+3) }
 \nonumber \\ &&
-
\ln(q\varpi_0/2)
\frac{z_0^5 w_0^5 q^6}{2^{6}}
\sum_{m=0}^{\infty}
\frac{(q/2)^{4 m}}{ \Gamma(4+2 m)}
\sum_{k=0}^{m}
\frac{ \varpi_0^{4 m-4k+2}}{\Gamma(2m-2k+2)}
\frac{(-1)^{2k}   z_0^{2 k} w_0^{2 k}
  }{\Gamma(k+1) \Gamma(k+3) }
 \nonumber \\ &&
+
\frac{q^4 z_0^5 w_0^5}{2^5}
\sum_{m=0}^{\infty}
\frac{(q/2)^{4m}}{\Gamma(2m+3)}
\sum_{k=0}^{m}
\frac{ \varpi_0^{4m-4k} }{\Gamma(2m-2k+1)}
\frac{(-1)^{2k}   z_0^{2 k} w_0^{2 k}  }{
\Gamma(k+1) \Gamma(k+3)}
\nonumber \\
&&\times
(\psi(2m-2k+1)+\psi(2m+3))
\nonumber \\
&&+
\frac{q^6 z_0^5 w_0^5}{2^7}
\sum_{m=0}^{\infty}
\frac{(q/2)^{4m}}{\Gamma(2m+4)}
\sum_{k=0}^{m}
\frac{ \varpi_0^{4m-4k+2}}{\Gamma(2m-2k+2)}
\frac{(-1)^{2k}  z_0^{2 k} w_0^{2 k}  }{
\Gamma(k+1) \Gamma(k+3)}
\nonumber \\
&&\times
(\psi(2m-2k+2)+\psi(2m+4))\,.
\nonumber \\
&=&
\sum_{m=0}^{\infty}
q^{4m}\left(
 T^{(1)}_m
+q^2 T^{(2)}_m
+q^4 T^{(3)}_m
+q^6 T^{(4)}_m
+q^4 T^{(5)}_m
+q^6T^{(6)}_m
\right)
\lab{eq:ksum}
\ea
where
\ba
T^{(1)}_m&=&
2^{-4m}
\frac{w_0^{2 m+5} z_0^{2 m+5}}{2\left(w_0^2+z_0^2\right)^2}
\frac{  \, _3F_2\left(1,1,\frac{3}{2};m+1,m+3;\frac{4 w_0^2
   z_0^2}{\left(w_0^2+z_0^2\right)^2}\right)}{ \Gamma (m+1) \Gamma (m+3) \Gamma (2 m+1)}\,,
\ea
and
\ba
T^{(2)}_m&=&
-2^{-4 m-2}
\frac{ w_0^{2 m+5} z_0^{2 m+5}}{2(w_0^2+z_0^2)}
\frac{  \, _3F_2\left(\frac{1}{2},1,1;m+1,m+3;\frac{4 w_0^2
   z_0^2}{\left(w_0^2+z_0^2\right)^2}\right)}{ \Gamma (m+1) \Gamma (m+3) \Gamma (2 m+2)}\,.
\ea
Moreover,
\ba
T^{(3)}_m&=&
-2^{-4 m-4}
\frac{(w_0 z_0)^{2 m+5}}{2\Gamma (2 m+3)^2}
\ln \left( \left(w_0^2+z_0^2\right)q^2/4\right)
   C_{2 m}^{(-2 m-2)}\left(\frac{w_0^2+z_0^2}{2z_0 w_0}\right)\,,
\ea
and
\ba
T^{(4)}_m&=&
2^{-4 m-6}
\frac{  (w_0 z_0)^{2 m+6} }{2\Gamma (2 m+4)^2}
 \ln \left( \left(w_0^2+z_0^2\right)q^2/4\right)
C_{2 m+1}^{(-2 m-3)}\left(\frac{w_0^2+z_0^2}{2 w_0 z_0}\right)\,,
\ea
where
$C^{(k)}_{2m}$ are Gegenbauer C polynomials.
The last terms use
\ba
T^{(5)}_m&=&
\sum_{k=0}^{m}
\frac{ (w_0^2+z_0^2)^{2m-2k} }{\Gamma(2m-2k+1)}
\frac{(-1)^{2k}   z_0^{2 k+5} w_0^{2 k+5}  }{
\Gamma(k+1) \Gamma(k+3)}
\frac{(\psi(2m-2k+1)+\psi(2m+3))}{2^5 \Gamma(2m+3)}\,,
\ea
and
\ba
T^{(6)}_m&=&
\sum_{k=0}^{m}
\frac{ (w_0^2+z_0^2)^{2m-2k+1}}{\Gamma(2m-2k+2)}
\frac{(-1)^{2k}  z_0^{2 k+5} w_0^{2 k+5}  }{
\Gamma(k+1) \Gamma(k+3)}
\frac{(\psi(2m-2k+2)+\psi(2m+4))
}{2^7 \Gamma(2m+4)}\,.
\ea

Performing the sum over $k$ in eqs.\ (\ref{eq:ksum}) and (\ref{eq:Tser})
one can calculate  the first few terms  of the expansion in $q$, i.e.~
\be
U_0=T^{(1)}_0=
\theta(z_0-w_0)
\frac1{4} z_0 w_0^5
+
\theta(w_0-z_0)
\frac1{4} w_0 z_0^5\,,
\ee
which determines the forward limit.
Similarly, 
\be
U_1=T^{(2)}_0=
\theta(z_0-w_0)
\frac{w_0^5 z_0}{48}(w_0^2-3 z_0^2 )
+
\theta(w_0-z_0)
\frac{w_0 z_0^5}{48}(z_0^2-3 w_0^2 )\,
\ee
reproduces the next-to-forward contribution which appears at order $q^2$.

Thus, summing all terms suppressed by $q^4$ one gets
\ba
U_2&=&T^{(1)}_1+T^{(3)}_0+T^{(5)}_0
=
\theta(z_0-w_0)
\frac{w_0^5 z_0 }{2^8 6}
\left(w_0^4-8 z_0^2 w_0^2
+18 z_0^4
-12 z_0^4 \ln \left(z_0^2 q^2 \e^{2 \gamma_E}/4\right)
\right)
\nonumber \\
&&+
\theta(w_0-z_0)
\frac{z_0^5 w_0 }{2^8 6}
\left(z_0^4-8 w_0^2 z_0^2
+ 18 w_0^4
-12 w_0^4 \ln \left(w_0^2 q^2 \e^{2 \gamma_E}/4\right)
\right)\,.  
\ea
Terms suppressed by $q^6$ read as follows
\ba
U_3&=&T^{(2)}_1+T^{(4)}_0+T^{(6)}_0
=
\theta(z_0-w_0)
\frac{ z_0 w_0^5 }{
2^{10} 90
}
\left(
w_0^6-15 z_0^2 w_0^4+90 z_0^4 w_0^2
+
170 z_0^6
\right.\nonumber \\ && \left.
-60 z_0^4 \left(z_0^2+w_0^2\right)
\ln   \left({z_0^2} q^2 \e^{2 \gamma_E} /4\right)
\right)
\nonumber \\
&
+
&
\theta(w_0-z_0)
\frac{w_0 z_0^5 }{
2^{10} 90
}
\left(
z_0^6-15 z_0^4 w_0^2 +90 w_0^4 z_0^2
+170 w_0^6
\right.\nonumber \\ && \left.
-60 w_0^4 \left(w_0^2+z_0^2\right)
\ln   \left(w_0^2 q^2\e^{2 \gamma_E}/4\right)
\right)\,.
\ea
Similarly, terms suppressed by $q^8$ give
\ba
U_4&=&T^{(1)}_2+T^{(3)}_1+T^{(5)}_1=
\theta(z_0-w_0)
\frac{
z_0 w_0^5}{2^{16} 3^3 5}
\left(w_0^8-24 z_0^2 w_0^6-270 z_0^4 w_0^4
+1420 z_0^6 w_0^2+645 z_0^8
\right.\nonumber \\ && \left.
-60 z_0^4 \left(3 z_0^4+8 w_0^2 z_0^2+3 w_0^4\right)
\ln   \left(z_0^2 q^2 \e^{2\gamma_E}/4\right)\right)
\nonumber \\
&&+
\theta(w_0-z_0)
\frac{
w_0 z_0^5}{2^{16} 3^3 5}
\left(z_0^8-24 w_0^2 z_0^6+270 w_0^4 z_0^4
+1420 w_0^6  z_0^2+645 w_0^8
\right.\nonumber \\ && \left.
-60 w_0^4 \left(3 w_0^4+8 z_0^2 w_0^2+3 z_0^4\right)
\ln   \left(w_0^2 q^2 \e^{2\gamma_E}/4\right)\right)\, . 
\ea


\end{document}